\documentclass{aa}  

\usepackage{subcaption}
\usepackage{float}
\usepackage{color}
\usepackage{natbib,twoopt}
\usepackage[hyphenbreaks]{breakurl}
\usepackage[breaklinks]{hyperref}  
                                      
\bibpunct{(}{)}{;}{a}{}{,}             
\definecolor{cobalt}{rgb}{0.06, 0.2, 0.65}
\hypersetup{
  colorlinks,
  citecolor=cobalt,
  linkcolor=[rgb]{0.8, 0.2, 1.0},
  urlcolor=cobalt,
}
\makeatletter
  \newcommandtwoopt{\citeads}[3][][]{\href{http://adsabs.harvard.edu/abs/#3}%
    {\def\hyper@linkstart##1##2{}%
     \let\hyper@linkend\@empty\citealp[#1][#2]{#3}}}
  \newcommandtwoopt{\citepads}[3][][]{\href{http://adsabs.harvard.edu/abs/#3}%
    {\def\hyper@linkstart##1##2{}%
     \let\hyper@linkend\@empty\citep[#1][#2]{#3}}}
  \newcommandtwoopt{\citealtads}[3][][]{\href{http://adsabs.harvard.edu/abs/#3}%
    {\def\hyper@linkstart##1##2{}%
     \let\hyper@linkend\@empty\citealt[#1][#2]{#3}}}
  \newcommandtwoopt{\citeyearads}[3][][]%
    {\href{http://adsabs.harvard.edu/abs/#3}
    {\def\hyper@linkstart##1##2{}%
     \let\hyper@linkend\@empty\citeyear[#1][#2]{#3}}}
\makeatother

\usepackage{graphicx}
\usepackage{txfonts}
\begin{document} 

   \title{AVISM: Algorithm for Void Identification in coSMology}

   \author{Óscar Monllor-Berbegal\inst{1}\thanks{oscar.monllor@uv.es},
          David Vallés-Pérez\inst{1,2},
          Susana Planelles\inst{1,3},
          Vicent Quilis\inst{1,3}
          }

   \institute{Departament d’Astronomia i Astrofísica, Universitat de València,
              46100 Burjassot (València), Spain
         \and
            Dipartimento di Fisica e Astronomia, Università di Bologna, Via Gobetti 93/2, IT-40129 Bologna, Italy
         \and
            Observatori Astron\`omic, Universitat de Val\`encia, E-46980 Paterna (València), Spain}

   \date{Received X XX, XXXX; accepted X XX, XXXX}

  \abstract
  {Cosmic voids are key elements in our understanding of the large-scale structure of the Universe. They are crucial to constrain cosmological parameters, understand the structure formation and evolution of our Universe, and they could also be pristine laboratories for studying galaxy formation without all the hassle due to environmental effects. Thus, the ability to accurately and consistently identify voids, both in numerical simulations and in observations, becomes mandatory.}
  {We present Algorithm for Void Identification in coSMology (\texttt{AVISM}), a new void finder for analysing both cosmological simulation outputs and observational galaxy catalogues. In the first case, the code should handle raw particle or cell data, dark matter halos or synthetic galaxy catalogues. 
  In the case of observational data, the code should be coupled with external tools providing with the required dynamical information to apply the algorithm. 
  This new numerical tool must be efficient in terms of computational resources, both wall time and memory.}
  {A set of numerical tests designed to assess the code's capabilities are carried out, including parameter robustness, computational performance and the use of the different matter components in a cosmological simulation. \texttt{AVISM}'s performance is also compared, both statistically and on a one-to-one basis, with the \texttt{DIVE} and \texttt{ZOBOV} state-of-the-art void finders using as input a dark matter halo catalogue from a large-volume cosmological simulation. An application to a galaxy survey is provided to demonstrate the code's ability to handle real data.}
  {We have designed a new void finder algorithm that combines geometrical and dynamical information to identify void regions plus a hierarchical merging process to reconstruct the whole 3D structure of the void. The outcome of this process is a void catalogue with complex boundaries without assuming a prior shape. This process can be repeated at different levels of resolution using finer grids, leading to a list of voids-in-voids and a proper description of void substructure.}
  {We present and release \texttt{AVISM}, a new publicly available void finder.}


   \titlerunning{Algorithm for Void Identification in coSMology}
   \authorrunning{Monllor-Berbegal, Vallés-Pérez, Planelles, Quilis}

  \maketitle
%

\section{Introduction}
\label{s:intro}

Cosmological voids are vast, nearly empty regions of the Universe that are sparsely populated by galaxies \citep{zeldovich1982giant} or any kind of matter and, hence, are underdense with respect to the background density at a given cosmological time. They arise from negative density perturbations in the initial fluctuation field \citep{sheth2004hierarchy} and their sizes span ranges from 10 to 20 $\mathrm{Mpc}/h$ or 20 to 50 $\mathrm{Mpc}/h$ \citep[e.g., see][where they report one of the largest known voids in the Universe]{kirshner1981million}, depending on the tracer used to define them \citep{van2011cosmic}. Although voids only account for $\sim15 \%$ of the mass of the Universe, they constitute $\sim80\%$ of its volume \citep{cautun2014evolution}, hence being much more prominent than any other structures surrounding them, such as filaments, walls or galaxy clusters.

Voids represent an excellent cosmic laboratory for studying the formation and evolution of galaxies in a medium mostly untouched by physical processes, like mergers, AGN activity, ram-pressure stripping, etc., that are present in high-density regions such as galaxy clusters or filaments. Thus, galaxies in voids are expected to evolve at a slower pace \citep{dominguez2023galaxies}, retaining the imprint of the early Universe \citep{van2011cosmic}. This results in different galactic properties (for instance, a less quenched stellar population) when compared to denser regions \citep[e.g., see][]{hoyle2012photometric, ricciardelli2014star, moorman2016star, rodriguez2024evolutionary, argudo2024morphologies}. For this reason, voids are well-suited for investigating galaxy formation as well as the impact of the large-scale structure (LSS) on the processes that drive galaxy evolution.

Furthermore, voids can also help to constrain cosmological parameters and, hence, to probe the $\Lambda$CDM ($\Lambda$-cold dark matter) cosmological model itself \citep{foster2009size, paz2023guess, contarini2024perspective, fernandez2024constraining, song2024void}. This is usually done by means of the excursion set formalism, first introduced by \citet{press1974formation} and later extended by \citet{epstein1983proto} and \citet{bond1991excursion}, which is a complete analytical description of the collapse and virialisation of overdense dark matter halos. The generalisation to voids, which is a similar but opposed problem, was later presented in \citet{sheth2004hierarchy}.

Besides void galaxies and cosmological parameter constraints, numerous contributions have been devoted to the study of the structure and evolution of voids. For instance, \citet{colberg2005voids}, \citet{ricciardelli2013structure, ricciardelli2014universality} or \citet{hamaus2014universal}, from the analysis of different cosmological simulations, presented universal profiles for the matter density inside voids and shed light on the evolution of void properties with cosmological time. Moreover, works like those by \citet{van1993voids} or  \citet{aragon2013hierarchical} described how, contrary to the common view, voids have complex internal structures and dynamics, having a hierarchical structure both in density and peculiar velocity fields, which results in the idea of 'voids-in-voids' or 'subvoids'. In fact, \citet{valles2021void} describe how simulated voids experience substantial mass inflows across cosmic history, suggesting that some of the gas present within voids originates from overdense regions such as filaments or clusters, challenging the idea of voids as pristine environments.

Despite the difficulty of defining a void and designing a method to identify empty regions, several algorithms have emerged to properly find and define these structures within galaxy surveys \citep[e.g.][]{foster2009size, pan2012cosmic} or cosmological simulations \citep[e.g.][]{ricciardelli2013structure} so as to study them. A first family of void finders would include those based on the watershed transform, first introduced in this context by \citet{platen2007cosmic} in the Watershed Void Finder (\texttt{WVF}), which identifies voids by treating the density field as a landscape, finding its basins. Technically, this algorithm is based on the Delaunay Triangulation Field Estimator (DTFE) \citep{bernardeau1996new,schaap2000continuous, van2008cosmic}. A similar approach was followed by \texttt{ZOBOV} \citep{neyrinck2008zobov}, which utilises the Voronoi Tessellation Field Estimator (VTFE) instead. Furthermore, \citet{sutter2015vide} proposed \texttt{VIDE}, a pipeline built around \texttt{ZOBOV} that, in addition, helps tracking voids throughout cosmic time with observational and simulated data. In this direction, another void finder following 
\texttt{ZOBOV}'S methodology is \texttt{REVOLVER}, described in \citet{nadathur2019beyond}. This family of tessellation algorithms is based primarily on geometrical arguments on the matter density field, assuming no shape for the void, which allows them to reconstruct any kind of structure.

A simpler methodology focuses on finding spherical regions with density contrast below a given threshold \citep[e.g.][]{kauffmann1991voids, hoyle2002voids, padilla2005spatial}, reason why they are called spherical void finders. Furthermore, a combination of both methods can be found in \citet{zhao2016dive}, where the authors describe \texttt{DIVE}, an algorithm involving Delaunay triangulation to efficiently compute the empty spheres constrained by a given discrete set of tracers (galaxies, dark matter particles, etc.). Both approaches impose spherical symmetry on the resulting void, which can be an issue if voids become more elongated as time progresses \citep{bos2012darkness}. However, they have the advantage of being able to connect to the void abundances developed in \citet{sheth2004hierarchy}. Besides, a natural extension of spherical void finders can be found in \citet{paz2023guess}, where they describe a novel void finder to capture more realistic, non-spherical void shapes, called 'popcorn voids'. The methodology involves recursively adding correction spheres to the initial spherical voids, providing a more accurate representation of the complex structures observed in cosmic voids. 

The aforementioned void finder families have been widely used in the literature and comparison projects have also been carried out \citep[e.g.][]{colberg2008aspen, cautun2018santiago, veyrat2023comparison}. Nevertheless, there is a third family of void finders that would involve considering not just the matter density field, but also dynamical information such as the peculiar velocity field. Because voids suffer super-Hubble expansion, that is, they expand at a faster rate than the rest of the Universe, they can be thought of as zones of positive velocity divergence, and algorithms can take advantage of this fact to find voids \citep[e.g.][]{lavaux2010precision, elyiv2015cosmic}.

In this work, we present and release \texttt{AVISM}, a new algorithm for void identification that results from a deeply revised and improved version of the void finder described in \citet{ricciardelli2013structure}, which uses both the density and velocity fields to find unstructured voids within the cosmic web. Therefore, this new void finder uses geometrical information but more importantly, also physical features to pinpoint empty regions in the Universe.

We extend the applicability of the code to survey and particle data and, hence, also to Smoothed-Particle Hydrodynamics (SPH) simulations. The original algorithm has been deeply revised in order to improve its efficiency and robustness, and at the same time, from the pure technical point of view, the code has been rewritten in order to gain a better performance, get a boost in its speed, as well as to be able to tackle large data volumes (e.g. in the case of simulations, more that $10^{10}$ particles).
For the sake of completeness, we present the comparison of \texttt{AVISM} with two of the most widely used void finders among the community, \texttt{DIVE} and \texttt{VIDE}, and an application to real data from the
\texttt{2M++} galaxy survey \citep{lavaux20112M++}.

The paper is structured as follows. In Sect. \ref{s:algorithm} we describe the algorithm and its methodology and characteristics, highlighting the changes and improvements with respect to the original version published in \citet{ricciardelli2013structure}. In Sect. \ref{s:mock_test} and \ref{s:comp_performance} we show the performance and scalability of the code when applied to an idealised test of several mock voids. In Sect.~\ref{sec:cosmological_simulations}, the algorithm is applied on two different state-of-the-art simulations to study the impact of different tracers on the final void distribution and also to display the substructure identification approach. Furthermore, we apply the algorithm along with two other state-of-the-art void finders to the halo catalogue from a cosmological simulation in Sect. \ref{s:comparison}. A detailed visual and statistical comparison of the results from the three methods is presented.  In Sect. \ref{s:survey} we provide two methodologies in which our code can handle galaxy survey data and we display the results when applied to the
\texttt{2M++} galaxy survey. Finally, in Sect. \ref{s:conclusions}, we summarise our work and discuss the main properties of our void finder. Appendix \ref{sec:appendix_test} provides details on the mock test construction, Appendix \ref{sec:appendix_excursionset} describes how we obtain the theoretical fit for the void size function and Appendix \ref{sec:appendix_metric} describes the approach followed to match different void catalogues.

\section{Algorithm}
\label{s:algorithm}

We present Algorithm for Void Identification in coSMology (\texttt{AVISM}), a new void finder approach that builds on the one described by \citet{ricciardelli2013structure}. 
The changes introduced in this new code can be grouped into two main categories. In the first one, new geometrical and dynamical conditions are considered to improve the accuracy of identification and classification of void regions. The second group of improvements are purely technical, with a great advance in efficiency and computational performance as a result of a deep rewriting of the main code routines in order to tackle the new era of cosmological simulations, which are increasingly more computationally demanding. The new algorithm is written in \texttt{Fortran} 2008 and efficiently parallelised using \texttt{OpenMP} directives. The code is publicly available in the corresponding GitHub repository\footnote{\url{https://github.com/oscarmonllor99/AVISM}}.

The new void finder can be applied either to outputs from cosmological simulations, halo catalogues or observational surveys, being able to work with the same level of accuracy and reliability in every case. When working on simulated data, either Lagrangian or Eulerian, the algorithm can identify voids using dark matter or gas, being able to tackle raw data from simulations including large numbers of particles or cells. Furthermore, it can treat a halo catalogue as a set of matter tracer particles to which the same algorithm can be applied to obtain voids. With a suitable density and velocity reconstruction method, the same procedure can also be straightforwardly applied to galaxy survey data.

\subsection{Input data}
\label{s:input}

One key feature distinguishing  \texttt{AVISM} with respect to other void finders in the literature is that it requires the velocity field to identify voids, since velocity divergence is essential to detect expanding regions and define their boundaries. Thus, the code is mainly based on the density $\rho$ and the velocity divergence $\nabla\cdot\mathbf{v}$ evaluated within a given cosmological volume. This data can originate directly from cosmological simulations, either in the form of a halo/galaxy catalogue or a full set of raw particles (or cells), or it can come from galaxy survey data. 

Originally designed to be coupled with the adaptive-mesh refinement (AMR) cosmological code \texttt{MASCLET}  \citep{quilis2004new, quilis2020cosmic}, this new version of our void finder can be run on any sort of format, being able to deal with large sets of particles (or cells) regardless of whether they stand for particles (dark matter or gas) or galaxies from a survey. To do so, our code needs to build a uniform auxiliary grid where the densities and velocity divergences are computed. This procedure has been achieved by implementing a method to transform a discrete particle distribution into a continuous distribution. This mechanism takes advantage of an SPH kernel in which the smoothing length is determined by a configurable parameter depending on the distance to the nearest neighbour particle (see details in Sect.~\ref{s:sph_interp}).

Periodic boundary conditions are also optionally supported by replicating the grid outside the input boundary limits. This is mandatory for cosmological simulations, where those boundary conditions are used to simulate the entire Universe in a limited volume.

\subsection{Continuous distribution from a discrete distribution}
\label{s:sph_interp}

As mentioned above, \texttt{AVISM} requires a set of physical quantities, namely the density and the velocity divergence, evaluated onto a grid.
In the case that the data comes from a grid-based cosmological simulation or from a real data catalogue previously processed with some software that translates these values on a grid, the void finder can directly read these data and be applied. 

When the data being analysed (either numerical or real) is composed of a collection of particles, an extra step is required to translate the discrete distribution of tracers into a continuous one onto a grid. This is one of the main changes implemented in the new version of our void finder, corresponding to a novel particle module which allows the interpolation of physical quantities described by a discrete particle distribution onto a grid. A complete and thorough description of this method can be found in \citet{valles2024vortex}. 

In our particular implementation, let us consider a set of particles whose positions, masses and velocities are known in a cubic region of side $L_0$. Inside this volume, we create a uniform grid with cells of size $\Delta x$. For assigning a continuous value of a physical quantity on the grid from a discrete set of tracers, we use a configurable parameter $N_\text{ngh}$ defining the number of neighbour particles contributing to each cell. By doing so, we can define two smoothing lengths. The first one is a smoothing length associated to each cell centre, defined as $h(\mathbf{x}) = \max \left( l_{N_\text{ngh}}, \Delta x \right)$, where $\mathbf{x}$ is the cell centre coordinates and
$l_{N_\text{ngh}}$ is the distance from the cell centre to the $N_\text{ngh}$-th nearest neighbour particle. On the other hand, for each particle $i$, we can introduce another smoothing length, $h_i$, defined as the distance to the furthest cell centre to which this particle contributes. Let us stress that although similar, the first one is associated with the cell centres, indicating the particles contributing to the quantity defined within a considered cell, whereas the second one is linked to particles describing the volume in which their quantities have to be spread out. 

With previous considerations in mind, it is possible to compute a continuous density field defined on the cell centres of the grid as:
\begin{equation}
\label{eq:dens_sum_1}
    \rho(\mathbf{x}) = \frac{1}{\Delta x^3}\sum_i m_i \widetilde{W}(|\mathbf{x}_i - \mathbf{x}|, h_i )\, ,
\end{equation}
where $m_i$ is the mass of particle $i$ and $\widetilde W$ is the SPH kernel properly normalised to guarantee mass conservation:
\begin{equation}
\label{eq:dens_sum_2}
    \widetilde{W}(|\mathbf{x}_i - \mathbf{x}|, h_i) = \frac{W(|\mathbf{x}_i - \mathbf{x}|, h_i)}{\sum_\text{cells}W(|\mathbf{x}_i - \mathbf{x}|, h_i)}\, .
\end{equation}
Here $W$ represents the kernel, which is set to the cubic spline kernel \citep[$\mathrm{M_4}$;][]{monaghan1985refined} by default in the code, although any other function can be easily supplied. The sums in Eq. \eqref{eq:dens_sum_1} and \eqref{eq:dens_sum_2} are taken over all particles ($\sum_i$) and all cell centres contributed by particle $i$ ($\sum_\text{cells}$), respectively. This procedure yields a conservative, continuous and differentiable density field without holes.

When reconstructing the peculiar velocity field, the strategy is slightly different. The velocity at the cell centres is computed using the volume-weighted contribution of their neighbouring particles: 
\begin{equation}
    \mathbf{v}(\mathbf{x}) = \frac{\sum_{i\in N_\text{ngh}} \frac{m_i} {\rho_i} \mathbf{v}_i W(|\mathbf{x} - \mathbf{x}_i|, h(\mathbf{x}))}{\sum_{i\in N_\text{ngh}}  \frac{m_i} {\rho_i} W(|\mathbf{x} - \mathbf{x}_i|, h(\mathbf{x}))}\, ,
\end{equation}
where $\mathbf{v}_i$ is the peculiar velocity of particle $i$ with mass $m_i$ and local density $\rho_i$. Here, the sum is performed for every cell with its $N_\text{ngh}$ nearest neighbours \footnote{We estimate the local density at the position of particle $i$ by summing the mass of all particles inside the sphere with radius equal to the distance to the $N_\text{ngh}$ neighbour, and dividing by the sphere volume.}. The continuous velocity field computed with this approach has the following characteristics:

\begin{enumerate}
    \item It is smooth and continuously differentiable, allowing us to correctly compute the velocity divergence.

    \item It does not leave cells for which no values are assigned, since we require every cell to be contributed by at least $N_{\text{ngh}}$ particles.

    \item The original information from a particle distribution is preserved as much as possible since the kernel shrinks in highly resolved zones. Besides, a volume-weighted approach is followed to properly describe the corresponding physical quantities inside voids, avoiding contamination from particles in denser zones.

    \item It is not conservative, since the volume integral of the continuised quantities does not match the integral volume of the original discrete distribution (unlike the density interpolation). 
    The discrepancy arises from the fact that, instead of performing a standard SPH summation  -- where each particle contributes based on its own smoothing length --,  we assign a kernel length to each cell centre to meet the requirements of our velocity assignment procedure. 
    Nonetheless, this issue is not relevant as the error is of $\sim 2\%$ for the $\mathrm{M}_4$ kernel with $N_\text{ngh}\approx50$ and it decreases with decreasing $N_\text{ngh}$ \citep{valles2024vortex}.
\end{enumerate}    

The search for neighbours inside a large collection of particles can be a very demanding issue. In order to keep the computational cost low, we have developed and implemented our own space-partitioning $k$-d tree algorithm \citep{bentley1975multidimensional} in \texttt{Fortran}, allowing seamless integration with our void finder. Besides, the tree construction is parallelised with \texttt{OpenMP} directives, further reducing the computational cost.

\texttt{AVISM} also allows the user to apply a Triangular Shape Cloud (TSC) kernel instead of the more complicated SPH procedure. This option is faster, conservative and it also produces continuous and differentiable fields. However, unless a coarse grid is used or a huge number of particles is considered, this method will leave cells with no values assigned (holes). 

A special case arises when the code is provided with data without the required velocity information to calculate its divergence. In this scenario, two options are contemplated in order to provide \texttt{AVISM} with such physical information.

A first approach to obtain the velocity field given the density distribution would be to use the expression for the continuity equation in the linear regime \citep{peebles2020principles}:
\begin{equation}
\label{div_estimation}
    \nabla \cdot \mathbf{v} = - f(t) a(t) H(t) \delta(\mathbf{x}, t)\, , 
\end{equation}
where $a(t)$, $H(t)$ and $\delta(\mathbf{x},t)$\footnote{We define the density contrast as $\delta = \frac{\rho}{\langle  \rho \rangle} -1$, with $\langle \rho \rangle$ being the mean density inside the input volume.} are the scale factor, the Hubble parameter and the density contrast at time $t$, and comoving coordinate $\mathbf{x}$, respectively. 
The perturbation parameter $f$ is well approximated by the expression $f = \frac{a}{\delta}\frac{d\delta}{da} \approx \Omega_m^{0.6}$. Although Eq.~\eqref{div_estimation} is obtained for the linear regime, its solution is a good approximation for a moderate non-linear regime \citep{van1993voids, hamaus2014universal}. Note, however, that in this case a restriction on $\nabla \cdot \mathbf{v} > 0$ does not carry any additional information to $\delta < 0$. Furthermore, in the special case of galaxy surveys, an additional step is also required to be applied to the density field in order to take into account galaxy bias \citep{kaiser1984spatial, cen1992galaxy} and redshift space distortions \citep[RSDs;][]{jackson1972critique, kaiser1987clustering}. This is why, in general, we would advocate for the usage of more advanced velocity field reconstruction methods before applying \texttt{AVISM}.

A more refined option would imply the usage of external tools, able to reconstruct the density and velocity fields taking into account the aforementioned issues. Several methodologies of this kind have been presented in the literature to extract the underlying density and velocity fields from galaxy surveys. Some of these tools generate linear reconstructions of the required fields \citep[e.g.][]{carrick2015cosmological, lilow2021constrained, ried2024velocity}, while more sophisticated options are able to obtain non-linear reconstructions \citep[e.g.][]{jasche2019physical, yu2019nonlinear, ganeshaiah2023large, mcalpine2025manticore}. For more details, we refer the reader to Sect.~\ref{s:survey}.

\subsection{Void-finding procedure}
\label{s:void_finding}

\begin{table*}
    \centering
    \caption{Summary of the main parameters used to run \texttt{AVISM}.}
    \begin{tabular}{p{5cm}p{1.5cm}p{2.5cm}p{8cm}}
    \hline
    \hline
    Parameter & Symbol & Value (default) & Description \\ \hline 
    \\
    \hspace{7 cm} Grid \\ \hline
    Minimum level & $\ell_{min}$ & 0 & - Coarse level for finding voids (only matters for grid-like inputs).\\
    Maximum level & $\ell_{max}$ & 0 & - Maximum level for finding voids.\\
    Coarse (minimum level) grid size & $N_x, N_y, N_z$ & - & - Number of cells in each direction for the grid corresponding to the coarse level.\\ 
    Comoving side of the box & $L_0$ & - & - Comoving side of the box in cMpc where the particles or data is placed.\\
    \hline
    \\
    \hspace{6 cm} \mbox{Particle interpolation} \\ \hline
    Number of neighbours & $N_\text{ngh}$& 32 & - 
  Number of neighbours that contribute to every cell in the density/velocity interpolation.  \\
    \hline
    \\
    \hspace{6.4 cm} \mbox{Void thresholds} \\ \hline
    Density contrast for void centres & $\delta_\text{1}$ & -0.6 & - Density contrast threshold applied to define cells that can grow and become voids. \\ 
    Density contrast for edges & $\delta_\text{2}$ & 10 & - Density contrast threshold to stop cube growing once the edge is found. \\ 
    Density gradient & $|\nabla\delta|_\text{th}$ & 0.25 $\mathrm{cMpc}^{-1}$  & - Density gradient threshold to stop cube growing once the gradient is too steep, close to the edges. \\ 
    Velocity divergence threshold & $\nabla \cdot \mathbf{v}_\text{th}$ & 0 $\; \mathrm{cMpc}^{-1} c$ & - Velocity divergence threshold to stop cubes growing into non-expanding regions.
    \\
    \hline 
    \hline
    \end{tabular}
    \tablefoot{For each parameter, its symbol, its default value (if applicable), and a brief description are provided. The values given to each parameter are justified in Sect.~\ref{s:void_finding}.}
    \label{tab1}
\end{table*}

Although most parts regarding the void-finding procedure implemented in \texttt{AVISM} have been rewritten, the core idea remains the same as in \citet{ricciardelli2013structure}. With $\rho(\mathbf{x})$ and $\mathbf{v}(\mathbf{x})$ defined at every cell centre $\mathbf{x}$ of a grid, the algorithm labels a cell as a candidate to be the centre of a void if the following criteria are fulfilled: i) the cell density contrast is below a specified threshold ($\delta < \delta_\text{1}$), and ii) it has a positive peculiar velocity divergence ($\nabla \cdot \mathbf{v} > 0$).

For every centre candidate, a cube is formed by extending the cell along the three Cartesian axes in both positive and negative directions. This growing procedure is repeated iteratively until one of the following conditions is met in any direction: 

\begin{itemize}
    \item Density gradient too steep ($|\nabla\delta| >|\nabla\delta|_\text{th}$), being $|\nabla\delta|_\text{th}$ a threshold value for the density contrast gradient.
    \item Velocity divergence above a given threshold (\mbox{$\nabla \cdot \mathbf{v}_\text{th}$})
    \item Density contrast above a given threshold (\mbox{$\delta > \delta_\text{2}$}), with $\delta_\text{2}$ being different from the density contrast threshold marking a tentative void centre ($\delta_\text{1}$).

\end{itemize}

This procedure yields a set of overlapping cubes, $\{C_i\}_{i=1}^{N_C}$, with $N_C$ the total number of cubes, covering all regions that are prone to being part of a void. It is important to note that a key change from the original void finder is the use of cubes instead of parallelepipeds, as we have tested that the combination of cubes of several sizes can better describe the geometry of voids. Thus, for each cube $C_i$ we tag all other cubes that are either overlapping or touching it, creating a list $\{C_{ij}\}_{j=1}^{N_\text{O}(i)}$ of related cubes that can be combined to obtain the complex 3D shape of voids, being $N_\text{O}(i)$ the total number of cubes overlapping or touching $C_i$.

In this direction, starting with the cube with the largest volume, $C_1$, the code initiates the first void, $V_1$, by merging to $C_1$ all cubes related to it
\begin{equation}
    V_1 = \bigcup_{j}^{} C_{1j}\, ,
\end{equation}
where the union is performed across all cubes overlapping or touching $C_1$. In the next step, we look for the second largest cube, $C_2$, which either could be found in two different situations:
\begin{enumerate}
    \item It is part of the $\{C_{1j}\}_{j=1}^{N_\text{O}(1)}$ list and, hence, already belongs to $V_1$. In this scenario, all cubes related with $C_{2}$ will be automatically added to $V_1$.
    
    \item It has not been merged yet and, hence, $C_2$ and all its related cubes will constitute a different void $V_2 = \cup_{j}^{} C_{2j}$. If any of the cubes associated to $C_2$ was already part of $V_1$, this particular cube cannot be included as a part of $V_2$. 
\end{enumerate}

We recursively apply this algorithm until all cubes $C_i$ are either the seed of a void or merged to an existing one. The outcome of this procedure is a sample of non-overlapping voids $\{V_k\}_{k=1}^{N_\text{voids}}$ that are built simultaneously inside (largest volume cubes) out (smallest volume cubes), with $N_\text{voids}$ the total number of voids. Furthermore, since this approach prevents a cube from being part of two voids, boundaries between them can be sharply obtained, preventing uncontrolled growth and complete percolation without the need to assume any prior on void shape. This is a major improvement with respect to the old version of this void finder \citep{ricciardelli2013structure}, where two user fixed parameters ($F_\text{min}$ and $F_\text{max}$) were needed to decide the minimal and maximal overlapping volume fraction in order to join or separate the void constituents (parallelepipeds in that case). On the other hand, when a region is shared between two cubes $C_i$ and $C_j$ belonging to different voids, the code solves the situation by assigning the overlapping volume to the biggest void.

\begin{figure}[h!]
\centering 
\includegraphics[width=0.6\linewidth]{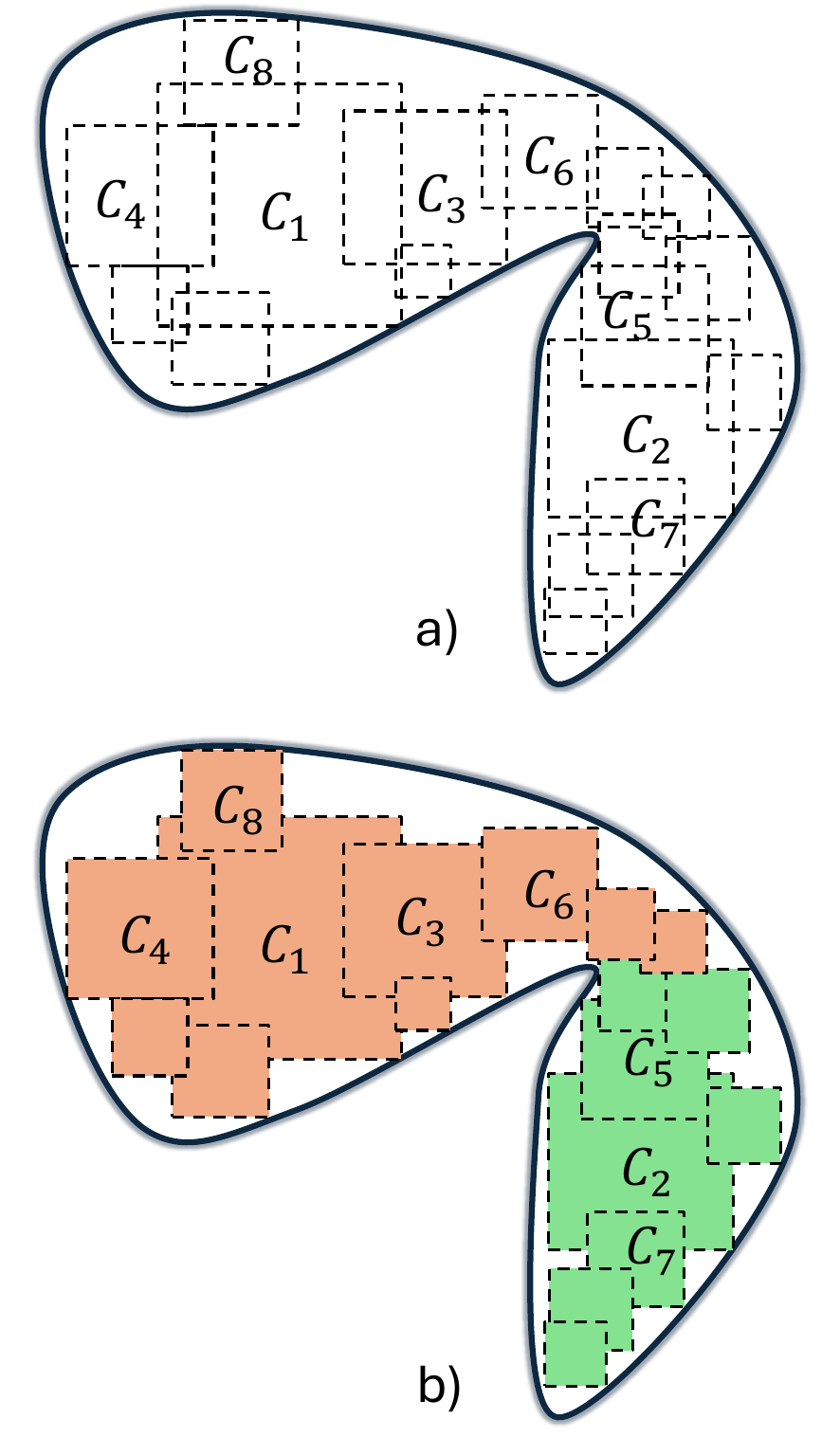}
\caption{Sketch of the void-finding procedure in an idealised 2D case. Top panel a) shows the complete set of volume-ordered cubes $\{C_i\}_{i=1}^{N_C}$ covering a region susceptible to be a void. Bottom panel b) displays how the algorithm is able to correctly group the cubes to produce well-separated voids illustrated in different colours.}

\label{fig:avism_algorithm}
\end{figure}

Figure \ref{fig:avism_algorithm} illustrates this procedure in an idealised 2D situation. The hierarchy of cubes, $\{C_i\}_{i=1}^{N_C}$, is displayed as squares of different surfaces (volumes in 3D). The first void, $V_1$, is initialised by considering the largest cube, $C_1$, and all the other cubes in contact or overlapping with it. The second largest cube is $C_2$ and, since it does not belong to $V_1$, a new void $V_2$ is created by merging $C_2$ with all its related cubes. The process runs on until no cubes are left to be assigned to a new or already existing void.

In order to avoid pathological situations, we have decided to extend every cube, $C_i$, by one cell along every axis (both positive and negative). In this manner, cubes that are not overlapping or touched but are very close neighbours can be linked together. After thorough testing on multiple grid resolutions, we determined that a one-cell extension leads to an optimal performance. 

In addition to previous steps, before delivering a final void catalogue, our method includes a post-processing algorithm ensuring that all voids become simply connected (without holes) by using the Breadth-First Search flood fill algorithm \citep[BFS;][]{cormen2022introduction}. This is mandatory, as steep density gradients or large matter concentrations can leave holes inside the 3D void structure. We have tested that the volume filling fraction before and after BFS changes little though, increasing by a small percentage ($1\%$ at most). 

\mbox{Table \ref{tab1}} summarizes the parameters used by \texttt{AVISM}. Taking as starting point the prescriptions given in \citet{ricciardelli2013structure}, by using a complete set of tests, the most crucial thresholds in the code have been set to:
\begin{itemize}
    \item{} $\delta_1 = -0.6$ is the density contrast threshold tagging cells as candidates to grow voids.
    \item{} $\delta_2 = 10$  is the density contrast threshold used to mark the void edge.
    \item{} $|\nabla\delta|_\text{th} =  0.25  \;\mathrm{cMpc}^{-1}$ is the density contrast gradient threshold that halts the growing of cubes by detecting the strong density gradient at the void boundaries. 
    \item{} $\nabla \cdot \mathbf{v}_\text{th} = 0 $ allows the detection of voids only in expanding regions. 
\end{itemize}

After a thorough experimentation with many parameter sets, it turns out that the values displayed in \mbox{Table \ref{tab1}} are a very robust choice for most applications.

\subsection{Void substructure}
\label{s:substructure}

The capability to disentangle void substructure and finding voids-in-voids is a crucial feature for any void finder in both simulations and observations.

\texttt{AVISM} can naturally tackle this problem by construction, as it is based on a hierarchy of nested grids at different levels of spatial resolution. This hierarchy starts from 
a coarse level, $\ell_{min}$, and reaches a given maximum level, $\ell_{max}$, with increasing resolution $\Delta x_{\ell} / \Delta x_{\ell+1} = 2$. The higher the level the better resolved are physical quantities and, therefore, their gradients and divergences become larger as a result of the sharper reconstruction of density and velocity field. 
Thus, by keeping fix the configurable parameters, those regions that a lower levels of the grid hierarchy have smoother density gradients and velocity divergence and, therefore, would satisfy the condition to belong to a void, now would be split into several sub-voids at higher levels of refinement.

From the algorithmic point of view, in order to identify substructures correctly, an extra condition has to be considered. When a cell is identified as a void centre candidate in a given level of refinement $\ell+1$, this cell will be immediately located within an already identified void at the lower level of refinement $\ell$. The process of growing and merging the cubes at level $\ell+1$  will be restricted to be inside the parent void at the lower level of refinement $\ell$ of the grid.

An example of substructure identification is presented in Sect.~\ref{s:substructure_iden}.

\section{Mock test}
\label{s:mock_test}
\begin{figure*}[h!]
\centering 
\includegraphics[width=0.85\linewidth]{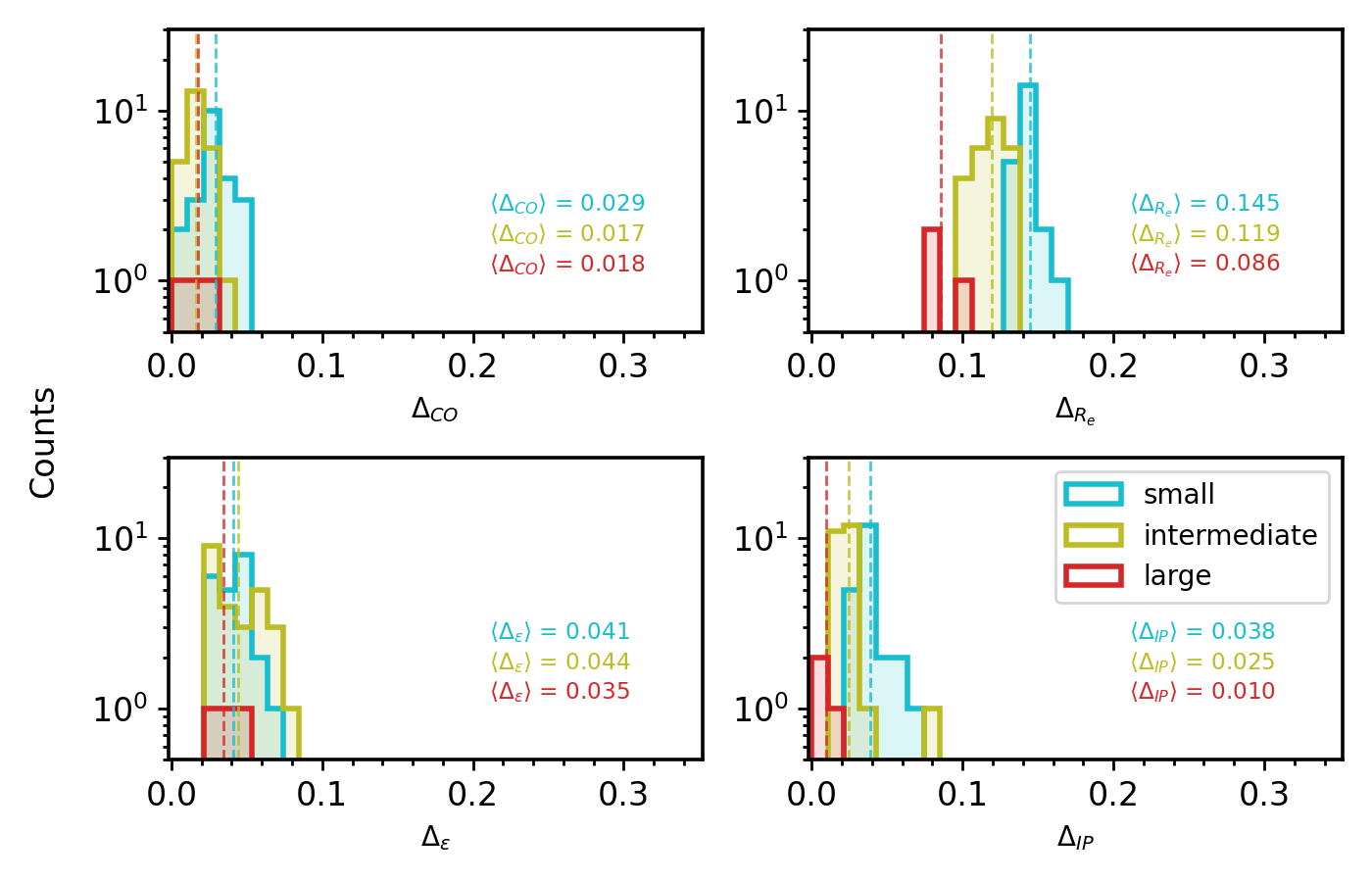}
\caption{
Relative errors between the mock void sample and the one obtained by \texttt{AVISM} as defined in the main text for four quantities: centre offset (top right), effective radius (top left), ellipticity (bottom left), and inverse porosity (bottom right). Colours stand for results for small (blue), intermediate (gold), and large (red) voids. The text within each panel displays the mean error of the considered quantity for each population.
}
\label{fig:test1_hist}
\end{figure*}

In order to justify the values adopted for the void thresholds in Table~\ref{tab1} and to assess the code robustness, we have built a test that consists of a particle-only non-periodic snapshot of $10^7$ particles in a \mbox{$L_0 = 147.5 \; \mathrm{Mpc}$}  \footnote{$100 \, h^{-1}\, \mathrm{Mpc}$ with $h=0.678$.} box at $z = 0$. Inside this box we have built $50$ ellipsoidal voids with a density profile as proposed in \citet{ricciardelli2013structure} (note that this profile is a generalisation of the one presented in \citealt{colberg2005voids}):
\begin{equation}
    \rho(<r) = \rho_e (r/R_e)^\alpha \exp\left([r/R_e]^\beta - 1\right)\,\, ,
\end{equation}
with $R_e$ the void effective radius\footnote{The void effective radius, $R_e$, is defined as the radius of the sphere with a volume equal to the actual void volume.}, $\alpha = 0.07$ and $\beta = 1.32$. These mock voids, which are not allowed to overlap, have a semimajor axis ranging from $12$ to $50$ Mpc. Following \citet{sheth2004hierarchy}, the mean density contrast inside $R_e$ is set to $\delta_e = - 0.8$. We refer the reader to Appendix \ref{sec:appendix_test} for more details on the construction of this test. 
The perfect elliptical shape of these voids represents a demanding challenge for \texttt{AVISM}, since its basic building blocks are cubes. However, the algorithm structure and the combination of multiple-sized cubes transform this apparent disadvantage into a powerful tool for describing complex void shapes.

This collection of idealised voids is a very stringent test as all void features are well-known and can be accurately computed. Thus, the comparison of different quantities estimated from the original sample (denoted by subindex $T$, standing for True) and from the counterparts produced by \texttt{AVISM} will shed light on the code's behaviour. In this particular application, we use a $128^3$ cell grid, which results in a resolution of $\sim 1 \;\mathrm{Mpc}$, and we set $N_\text{ngh} = 32$ (the default value). For the sake of clarity, and in order to study the code's performance depending on voids' size, we segregate voids into three sizes: small ($R_e < 10 \; \mbox{Mpc}$), intermediate ($10 \; \mbox{Mpc} < R_e < 17 \; \mbox{Mpc}$) and large ($R_e > 17 \; \mbox{Mpc}$). 

Figure \ref{fig:test1_hist} displays clockwise the relative errors, defined as $\Delta_X = |1-X/X_T|$, for four quantities: the centre offset, the effective radius, the inverse porosity\footnote{We define the inverse porosity as $\mathrm{IP} = V_E/V$ \citep{shandarin2006shapes}, where $V_E$ is the volume of the ellipsoid fitting the void and $V$ is the actual void volume.}, and the ellipticity\footnote{We define the ellipticity of a void as $\varepsilon = 1 - c/a$, where $a$ and $c$ are, respectively, the lengths of the major and minor axis of the associated ellipsoid.}.
The results for small, intermediate, and large voids are shown in blue, gold, and red, respectively.

As expected, the larger the void, the better the void finder is able to reproduce the true values of its properties, since there are more resolution elements (grid cells) to catch the true shape. 
In contrast, in small voids, a single cell can be a $\sim 10 \%$ of the effective radius, thus producing larger uncertainties in the void properties, especially in the size (effective radius) determination. Intermediate voids are halfway between the other two behaviours, hence obtaining a smooth transition between the different accuracies.

Naively, one could think of using finer grids to overcome possible resolution effects, but depending on the data, this could worsen the situation. For instance, using a $512^3$ grid on the mock test input, many cells could be left without particles, leading to over-smoothed and noisy data after the interpolation process. On the contrary, increasing the number of particles always improves performance, as more numerical elements are used to sample the underlying density and velocity fields.

The resolution of the grid should also be chosen depending on the size of the volume under study and the application. When searching for voids at $z \approx 0$, it could be counterproductive to resolve regions smaller than $1-2 \; \mathrm{Mpc}$, since this is the realm of galaxy clusters and filaments. Therefore, too high numerical resolutions could lead to the creation of undesired boundaries that spuriously split voids. We consider a cell size of $\mathrm{1-2 \; Mpc}$ a proper resolution to find large voids in the coarse level of resolution. When considering a hierarchy of nested grids with increasingly higher numerical resolution, as discussed before, the higher resolution will capture steeper gradients that naturally divide voids into smaller parts, thus creating a structure of voids-in-voids. 

In summary, \texttt{AVISM} is able to properly recover the $50$ mock voids inside the analysed volume and correctly reproduce their main properties, especially for $R_e > 10 \; \mathrm{Mpc}$. Details on computational performance can be found in Sect.~\ref{s:comp_performance}.

\section{Application to cosmological simulations}
\label{sec:cosmological_simulations}

After applying the new void finder to an idealised mock test, in this section we analyse the outputs from two complex cosmological simulations of a very different nature. Besides studying different aspects of the performance of the code, we illustrate how \texttt{AVISM} can handle such different inputs.

In the first case, we use snapshots from a moving-mesh code (Lagrangian approach) in order to assess how the use of different numerical tracers, namely, dark matter particles, gas particles, dark matter halos or galaxies, can affect to the void identification process. Moreover, this particular simulation uses a large number of dark matter and gas particles, thus emphasising the ability of the code to deal with a large number of numerical tracers.

In a second application, we analyse a grid-based simulation (Eulerian approach) to show an example of substructure identification.

\subsection{Numerical tracers}
\label{sec:TNG_tracers}

\begin{figure*}[h!]
\centering 
\includegraphics[width=0.9\linewidth]{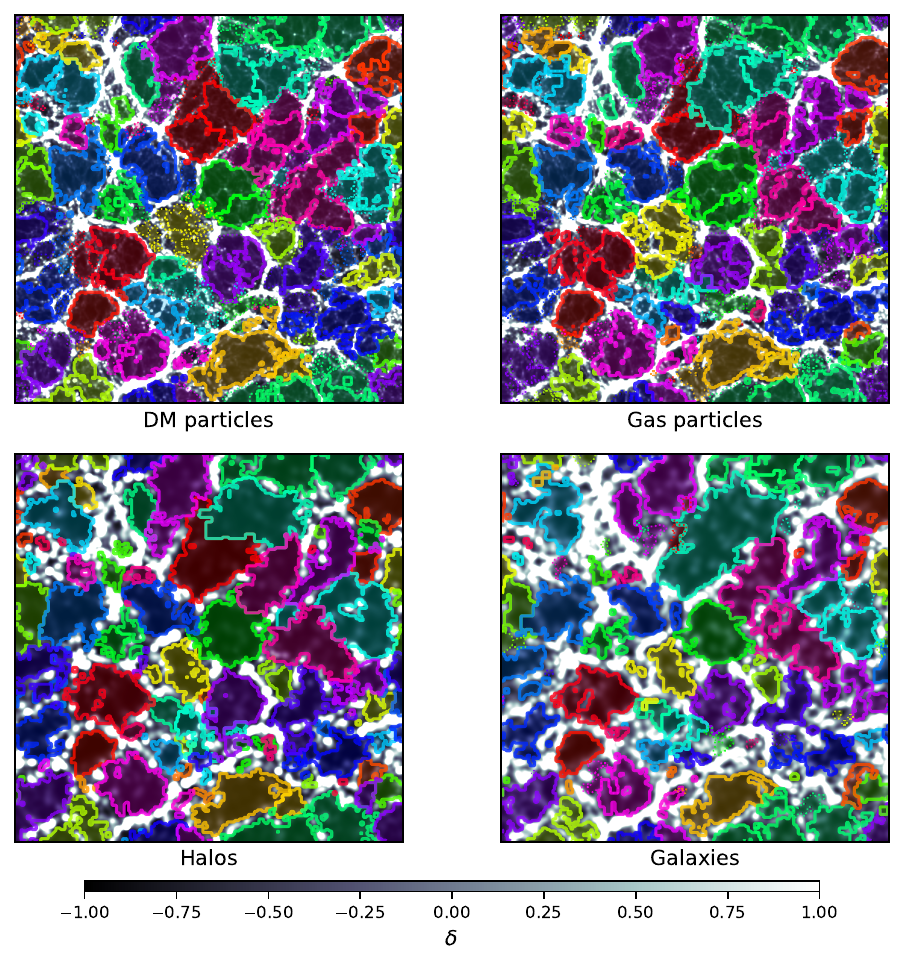}
\caption{Distribution of voids obtained by \texttt{AVISM} when applied to a $z = 0$ snapshot of the \texttt{TNG300-2} simulation on four different matter tracers: dark matter particles, gas particles, halos and galaxies. The images show, for each tracer, all voids intersecting a thin slice of 302.6 $\mathrm{Mpc}$ side and $\approx 10 \; \mathrm{Mpc}$ depth, together with the integrated density field, for which a colour-bar is displayed. Voids matching another from the reference catalogue (using halos as matter tracers) with DSC coefficient larger (smaller) than 0.4 are displayed using the same colour and continuous (dotted) lines.}
\label{fig:TNG_demo}
\end{figure*}

A crucial aspect underpinning the void-finding problem refers to the numerical elements used to define the physical quantities that, in turn, are used to identify voids. When analysing cosmological simulations, different flavours of numerical tracers can be used: dark matter particles, gas particles (or cells), halos, or galaxies. One could have a legitimate concern about how the chosen tracer affects the outcome of the void finder.

In order to demonstrate \texttt{AVISM} capabilities to handle different matter tracers, and their effect on the void identification, we have run the void finder over a $z = 0$ snapshot of the \texttt{TNG300-2} cosmological simulation, from the \texttt{IllustrisTNG} suite \citep{nelson2019illustristng}. This simulation models the evolution, from redshift $z = 127$ to $z = 0$, of a cubic volume with $302.6\;\mathrm{cMpc}$ side length, containing $1250^3$ gas and dark matter particles. It incorporates a comprehensive galaxy formation model that accurately tracks the formation and evolution of galaxies over cosmic time
 \citep{weinberger2016simulating, pillepich2018simulating}. The void finding algorithm has been applied with the default values described in Sect. \ref{s:void_finding} to four different tracers: all dark matter particles, all gas particles, all halos and all galaxies. The results are shown in Fig. \ref{fig:TNG_demo}, where we display the distribution of voids inside a thin slice of 302.6 $\mathrm{Mpc}$ side length and $\approx 10 \; \mathrm{Mpc}$ depth, together with the integrated density contrast field.
 
To provide a more intuitive comparison among the four void catalogues produced using the four different tracers, we use the Dice-Sørensen coefficient (DSC) (see Appendix~\ref{sec:appendix_metric} for details) as a metric to measure the degree of matching. We take as the reference catalogue the one produced by the dark matter halos, being a compromise between the number of numerical tracers. Voids that in the other three catalogues match with other void from the reference catalogue with a DSC higher than 0.4 are displayed by a continuos contour line with the same colour. Those would be matches that have a high volume intersection. In the same manner, void matches that have a DSC value smaller than  0.4 are considered a likely counterpart, although their intersected volume would be smaller. They are also drawn using the same colour but with dotted contour lines.

As a general trend, there is a reasonable match among the outcomes produced by the four different tracers. Nevertheless, the use of a larger number of numerical tracers leads to different spatial distributions of the physical quantities, with sharper features that would produce some voids in the reference catalogue to be split into smaller ones in the dark matter or gas particles catalogues. In the same line, the lesser tracer particles used, the smoother the density and velocity fields. This is the reason why the catalogue based on galaxies has the tendency to produce larger voids.

\begin{figure}[h!]
\centering 
\includegraphics[width=1.\linewidth]{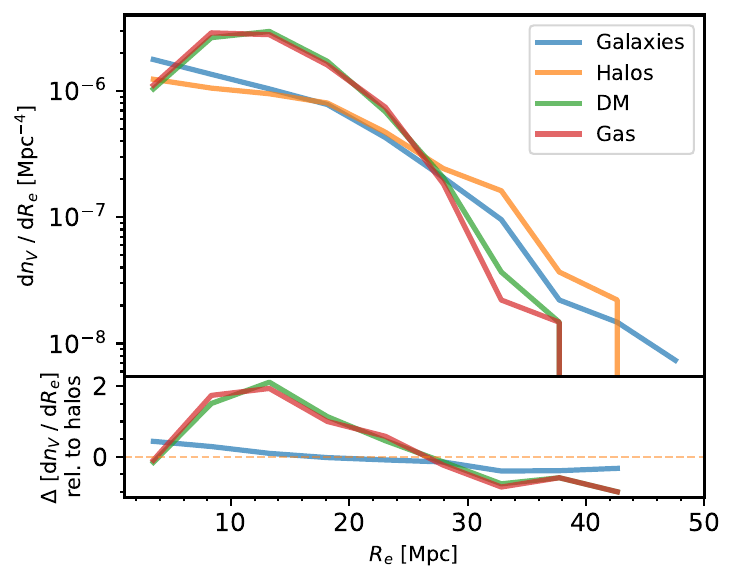}
\caption{
Top panel: void size function for the different void catalogues obtained by \texttt{AVISM} when run on four different numerical tracers of the \texttt{TNG300-2} cosmological simulation from the \texttt{IllustrisTNG} suite. Bottom panel: relative difference with respect to the reference void size function obtained using halos as tracers.
}
\label{fig:statistics_tracers}
\end{figure}

A more complete perspective is given by the void size function (VSF)\footnote{The void size function is defined as the number density of voids per effective radius.} presented in Fig. \ref{fig:statistics_tracers}, where the void catalogues produced by the four different numerical tracers are analysed. 
Two distinct behaviours are obtained: halos and galaxies tend to yield larger voids, while dark matter and gas particles tend to produce smaller ones. These results reinforce the previously stated idea that the number of considered numerical tracers directly impacts the level of detected substructure.

A remarkable result is the fact that, although voids sizes and locations can vary to some extent, statistically, the results produced by the algorithm present an outstanding robustness against huge variations in the number of numerical tracers ($\sim 10^9$ for gas or dark matter particles and $\sim 10^6$ for halos or galaxies).
 
\subsection{Substructure identification}
\label{s:substructure_iden}

In order to provide an example displaying the outcome of our substructure approach, we apply \texttt{AVISM} on a snapshot at $z \approx 0$ from a simulation produced by the \texttt{MASCLET} hydrodynamic and $N$-body code \citep{quilis2004new}, which is based on an adaptive-mesh refinement (AMR) scheme. This simulation describes the evolution of a $100 \; \mathrm{cMpc}/h$ cosmological box using nine levels of refinement, which allows a peak spatial resolution of $1.1 \; \mathrm{ckpc}$. It is similar to the one described and applied in \citet{ricciardelli2013structure} to study void structures, but with a better spatial resolution. The grid refining criteria are chosen to ensure a proper description of the physical quantities in void regions and, hence, to obtain a proper evolution of these structures in the simulated volume.

\begin{figure}[h!]
\centering 
\includegraphics[width=1.\linewidth]{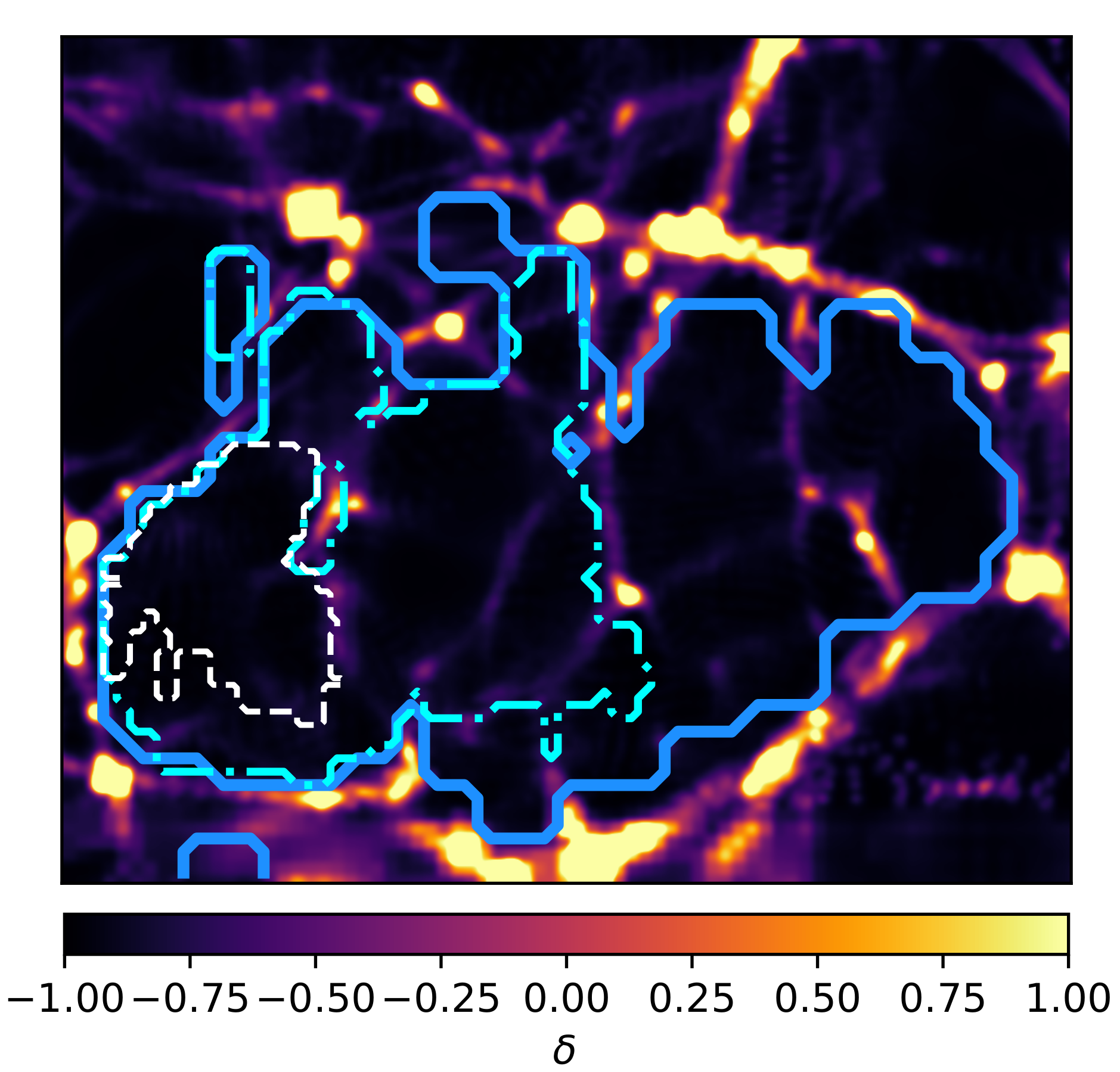}
\caption{Slice of a zoom in on a region centred at a $R_e \approx 40 \; \mathrm{Mpc}$ void at $\ell = 0$ (dark blue solid line) together with its biggest sub-void at $\ell = 1$ (light blue dash-dotted line) and a substructure of that sub-void at $\ell = 2$ (white dashed line). The slice is $5 \; \mathrm{Mpc}$ depth. The colour palette displays the integrated density contrast. The analysis was performed on a snapshot of a \texttt{MASCLET} simulation at $z = 0$. More substructures are obtained for the same void and its sub-voids; however, only one of each kind is shown for the sake of clarity.}
\label{fig:subvoids}
\end{figure}

Regarding the void-finding methodology, the values for the thresholds used to obtain these results correspond to the default configuration. In addition, in this case, the void finder is run with a three-level grid hierarchy ($\ell = 0,1,2$) from which substructure can be studied in detail. Fig.~\ref{fig:subvoids} shows a slice of $5 \; \mathrm{cMpc}$ depth zooming in on a $R_e \approx 40 \; \mathrm{Mpc}$ void at $\ell = 0$ (dark blue, solid line) together with its biggest sub-void at $\ell = 1$ (light blue, dash-dotted line) and a substructure of that sub-void at $\ell = 2$ (white dashed line). Note that, in order for the illustration to be clear, we are only showing a void at each level of the hierarchy, but more substructures were obtained for the $\ell=0$ void in the other two levels. As explained above, the same thresholds are applied for the three different resolutions. At $\ell = 0$ (lowest resolution), the physical quantities are smooth and, hence, the velocity divergence or density gradient do not present substantial variations in space, hence obtaining larger voids. For $\ell > 0$, the increase in resolution makes the divergences and gradients steeper. Consequently, more cells exceed the thresholds to stop void growth, yielding a set of smaller voids that are contained inside the larger ones at lower levels of the hierarchy and can be understood as physical substructures. Let us draw attention to how less dense filaments and tendrils become the boundaries of the sub-voids at higher levels of refinement.

\section{Computational performance}
\label{s:comp_performance}

\begin{figure}[h!]
\centering 
\includegraphics[width=1\linewidth]{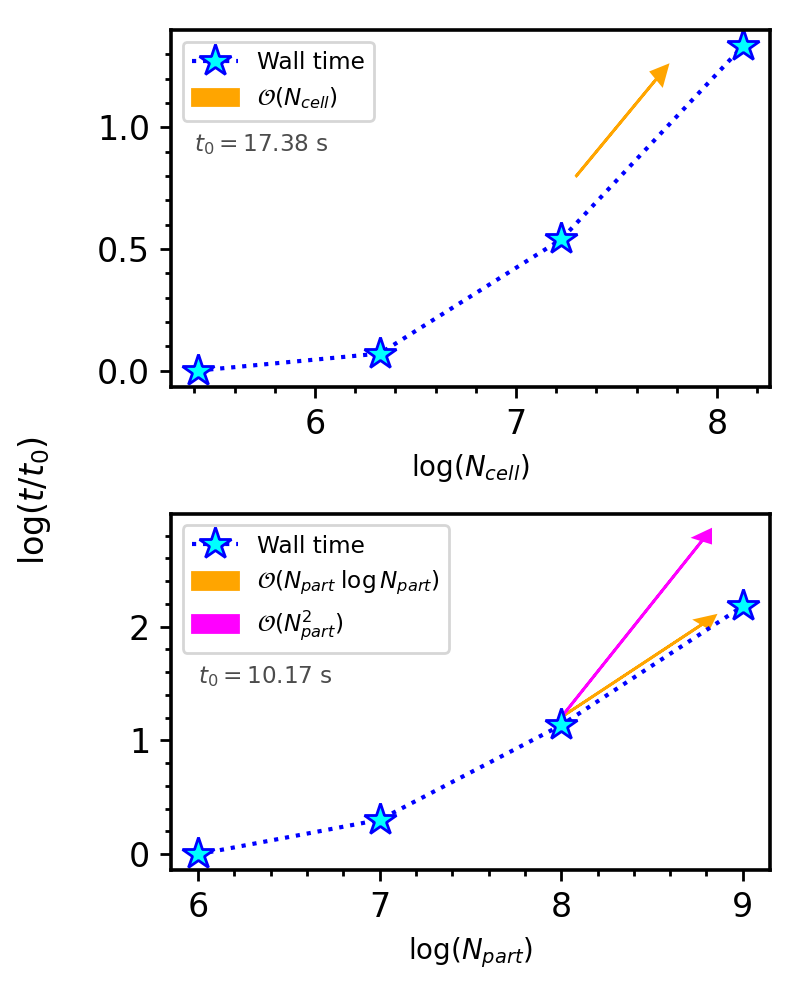}
\caption{This figure shows the code time complexity. Times are normalised with respect to $t_0$, the wall time for the minimum number of cells or particles considered in this test, which is shown below both panel legends. Top panel: wall time against the number of cells. Bottom panel: wall time against the number of particles that have to be interpolated onto the grid. In yellow and purple different time complexities are given as a reference. Logarithms are taken in base $10$.}
\label{fig:scaling}
\end{figure}

\begin{figure}[h!]
\centering 
\includegraphics[width=1\linewidth]{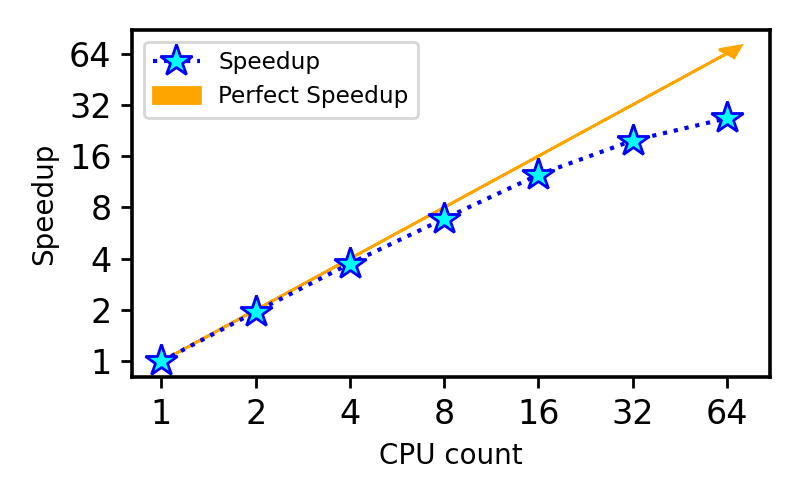}
\caption{\texttt{AVISM}'s speedup against the number of CPU cores used. In yellow, a perfect speedup is given as a reference.}
\label{fig:cpu}
\end{figure}

In order to evaluate the code's computational performance, we have applied \texttt{AVISM} to the mock test volume described in Sect. \ref{s:mock_test} using different grid and CPU configurations. In addition, in order to assess the particle-to-grid interpolation scalability, we have also produced different versions of the test varying the number of particles. The code was compiled by the GNU Fortran 11.4 compiler and was run on an AMD Ryzen Threadripper PRO 3995WX (64 cores) CPU.

Building a $k$-d tree implies an initial cost of $\mathcal{O}(N_\text{part} \log N_\text{part})$ \citep{bentley1975multidimensional}, with $N_\text{part}$ the number of input particles. Then, searching for the neighbours around some point, implies a $\mathcal{O}(\log N_\text{part})$ complexity. Hence, when using a grid consisting of $N_\text{cell}$ cells, the particle interpolation process scales as $\mathcal{O}( N_\text{cell} \log N_\text{part})$ for the velocity, and as $\mathcal{O}( N_\text{part} \log N_\text{part})$ for the density. 

Regarding the void-finding algorithm (see Sect.~\ref{s:sph_interp}), it requires the creation of a set of cubes covering all regions susceptible to belonging to a void. To achieve this, it loops over all cells belonging to the grid, growing a cube where the corresponding physical thresholds are fulfilled. Thus, since not every cell has to be expanded (not every cell fulfils the necessary physical thresholds), and many will already be part of a cube, we expect the algorithm to have, at maximum, a $\mathcal{O}(N_\text{cell})$ complexity. Once all cubes are created, they are merged depending on whether they overlap or touch each other, leading to a complexity of \(\mathcal{O}(N_{C}^2)\), with $N_C$ the total number of cubes. Nevertheless, we leverage our particular implementation of the $k$-d tree algorithm, allowing us to accelerate the process by restricting the merging procedure to cubes that are within a certain distance. Time complexity thus becomes $\mathcal{O}(N_{C} \log N_{C})$, but, since $N_\text{C} \propto N_\text{cell}$, the merging process ultimately has a $\mathcal{O}(N_{\text{cell}} \log N_{\text{cell}})$ complexity, at most. On the other hand, as explained in Sect.~\ref{s:void_finding}, the final step is to get rid of possible holes in the final void structures by applying the BFS method, which presents $\mathcal{O}(N_\text{cell})$ scaling.

The volume of the region we want to analyse, either from simulations or observations, is also a key ingredient affecting the performance of the void finder.  As the Universe's volume is mostly occupied by voids, the number of these structures will increase almost proportionally to the rise of the volume of the considered region. Besides, the number of non-void structures, such as clusters, filaments and sheets also increases, making more costly the process of identifying the cells in voids, the creation of cubes, and their mergers to produce the final voids.  

In order to check the time complexity in \texttt{AVISM}, we have performed two different sets of runs of the mock test presented in Sect.~\ref{s:mock_test}. In the first, we fix the number of particles to $N_\text{part} = 10^7$ and vary the grid number of cells from $64^3$ to $512^3$ in powers of $2^3$. Then, we perform a second test fixing the grid to $128^3$ and vary the number of particles from $10^6$ to $10^9$ in powers of $10$. Both tests have been run using 16 CPU cores. The results can be found in Fig.~\ref{fig:scaling}. The top panel shows  how the wall time scales as $\mathcal{O}(N_\text{cell})$ at maximum,  better than previous expectations. Regarding the number of particles, the bottom panel also exhibits a closer time complexity to the expected $\mathcal{O}(N_\text{part} \log N_\text{part})$. 

Regarding the code scalability when running on parallel systems, the speedup of the current version is good, although some parts of the code cannot be parallelised and, therefore, result in bottlenecks holding the scalability. The void expansion and merging processes present some problematic race conditions and most parts of these code sections have to be run serially. Nevertheless, the particle-to-grid interpolation (including $k$-d tree construction) which represents, depending on the run, the most computationally expensive part, can be perfectly parallelised by means of \texttt{OpenMP} directives. The speedup for the mock test described in Sect.~\ref{s:mock_test}, using a single grid level $\ell = 0$ of $256^3$ cells and $10^7$ particles, is presented in Fig.~\ref{fig:cpu}. The computational time decreases significantly up to $32-64$ cores, after which the speed-up starts to flatten out. Indeed, this scaling occurs due to the fact than an increasing number of threads cannot reduce the computational cost of the void-finding and merging processes, as these are run serially. This figure shows a balanced case in which the number of particles and cells are within a similar order of magnitude ($\sim 10^7$). In unbalanced cases the situation can get better (worse) if the number of particles is significantly greater (lower) than the number of cells in the grid.

Let us stress one final feature of \texttt{AVISM} regarding its computational performance. The code can handle in an extremely efficient way, both in terms of memory and CPU time, very large volume datasets. As a particular example to highlight this point,  the \texttt{TNG300-2} simulation snapshot at $z = 0$ was analysed using all dark matter particles ($1250^3$) with the default set of parameters described in Sect. \ref{s:void_finding}. Running the code with 32 cores took 1 hour and 10 minutes and allocated a maximum of $\sim 360$ GB of RAM at its consumption peak.

\section{Comparison with DIVE and ZOBOV}
\label{s:comparison}

Every algorithm has its own strengths and caveats and, therefore, when describing a new computational tool is crucial to contextualize its performance by comparing with some of the codes widely used by the community. In this sense, it is of the utmost importance to compare \texttt{AVISM} with some of the most popular void finders. To carry out this comparison, we have considered two well-known and widely used codes, each of them broadly representing the two most common approaches used in the void-finding algorithms:

\begin{itemize}
    \item \texttt{DIVE} \citep[][]{zhao2016dive}: Delaunay trIangulation Void findEr is a C++ tool for identifying all empty spheres that are constrained by four elements of a point set, using the Delaunay Triangulation (DT) technique. It is able to resolve all the maximal spheres that are empty of whatever element that is used as tracer, such as galaxies in either a real survey volume or a periodic simulation box. These spheres are regarded as a special type of cosmic voids (DT voids) which are allowed to overlap with each other. The output of the code are the spatial positions of the centres of these spheres, along with their radii. However, these spheres are not actual voids but rather candidates for being voids, since overlaps have to be eliminated, obtaining a set of disjoint voids. Despite that \texttt{DIVE} was conceived for finding large-scale underdensities in the very diluted sample of luminous red galaxies (LRGs), not for studying void structure and dynamics, it has been widely used recently \citep{contarini2022euclid, tamone2023void, fernandez2024constraining} to study void statistics and constrain cosmological parameters. Caution must be taken, however, when comparing it with other void finders due to its particular approach.  With this in mind, our goal is to compare \texttt{AVISM}'s voids with those of \texttt{DIVE} on a simple void-placement and size basis.

    \item \texttt{VIDE} (\texttt{ZOBOV}) \citep[][]{sutter2015vide}: Void IDentification and Examination toolkit is an open-source Python/C++ code for finding cosmic voids in galaxy redshift surveys and $N$-body simulations. It is built on \texttt{ZOBOV} \citep{neyrinck2008zobov}, which builds a Voronoi tessellation of the tracer particle population and utilises the watershed transform to group Voronoi cells into zones, eventually identifying voids. \texttt{VIDE} has several modifications and improvements with respect to \texttt{ZOBOV}, both in terms of computational performance and the algorithm's design. The outcome of this void finder is extensive. We focus on the void volume, particles belonging to each void and volume occupied by each particle's Voronoi cell. This void finder targets the same goals as \texttt{AVISM}, namely the 
    study of the full void structures and substructures across cosmic time in simulated and real data. In order to make the comparison clearer, we focus on void placement and sizing, such as in the \texttt{DIVE}. Moreover, throughout the rest of the paper, we refer to \texttt{VIDE} as \texttt{ZOBOV}, for the sake of clarity.

\end{itemize}

In order to compare the performance of \texttt{AVISM} with that of \texttt{DIVE} and \texttt{ZOBOV}, we will apply the three void finders to the same simulation output. For the sake of a complete comparison, the simulation has to satisfy the following requirements:

\begin{enumerate}
    \item It is publicly available, for the sake of reproducibility.
    \item It describes a large cosmological volume ($L_0\gtrsim400 \; \mathrm{Mpc}$), thus containing sufficient void statistics.
    \item It already has an available halo catalogue to which we can apply the void finders.
\end{enumerate}

The comparison is carried out using a halo catalogue as an input, first, because it is generally faster for all void finders since there are less tracers to process and, second, because \texttt{DIVE} is particularly designed for this kind of input or survey data (low density of tracers). A suitable simulation fulfilling all of these requirements is \texttt{mini-UCHUU}, from the \texttt{UCHUU} $N$-body simulations suite \citep{ishiyama2021uchuu}\footnote{\url{https://skiesanduniverses.org/Simulations/Uchuu/}}. It uses Planck cosmology \citep{aghanim2020planck} with $\Omega_m = 0.3089$, $h = 0.6774$, $\sigma_8 = 0.8159$ and $n_s = 0.9667$ with a cosmological box of $L_0 = 400 \; \mathrm{Mpc}/h \approx 591 \; \mathrm{Mpc}$ at $z = 0$ containing $2560^3$ particles with a softening length of $\varepsilon = 4.27 \; \mathrm{Mpc}/h$. All simulation outputs have already been analysed by means of the \texttt{ROCKSTAR} halo finder \citep{behroozi2012rockstar}. We are interested in the last output ($z = 0$), where there are $N_h \approx 1.7 \times10^{7}$ halos with $\mathrm{M}_\text{vir} > 10^{10} \; \mathrm{M}_\odot$. While \texttt{DIVE} only requires the position of each tracer, \texttt{ZOBOV} also needs mass and \texttt{AVISM} needs positions, velocities and masses. Moreover, the \texttt{DIVE} and \texttt{ZOBOV} inputs have been reduced in order to accommodate the number of tracers to the requirements of each code. The outcome of the void-finding processes for these algorithms, unlike \texttt{AVISM}, strongly depends on the density of tracers (e.g., see \citealt{massara2022velocity}). 
Hence, we properly choose this quantity in order the void finders to produce void samples with good statistical properties \citep[][]{sutter2014sparse}. Thus, following the analysis performed in \citet{zhao2016dive} and the approach considered in \citet{fernandez2024constraining}, we will only use halos above $\mathrm{M}_\text{vir} > 10^{13} \; \mathrm{M}_\odot$ in order to achieve a tracer density of $\overline{n} \approx 5 \times10^{-4} \; (\mathrm{Mpc}/h)^{-3}$ for the \texttt{DIVE} case. Regarding \texttt{ZOBOV}, using a mass cut of $\mathrm{M}_\text{vir} > 10^{12} \; \mathrm{M}_\odot$ we obtain a tracer density of $\overline{n} \approx 5 \times10^{-3} \; (\mathrm{Mpc}/h)^{-3}$.
Hence, to perform a comparison as fair as possible,  \texttt{DIVE} and \texttt{ZOBOV} will be applied to a subset of the provided input, whereas \texttt{AVISM} will be run on the whole input, using a single level grid (substructure will be ignored in this comparison) of $256^3$ cells together with the default thresholds described in Sect.~\ref{s:void_finding}.

\begin{figure}[h!]
\centering 
\includegraphics[width=1\linewidth]{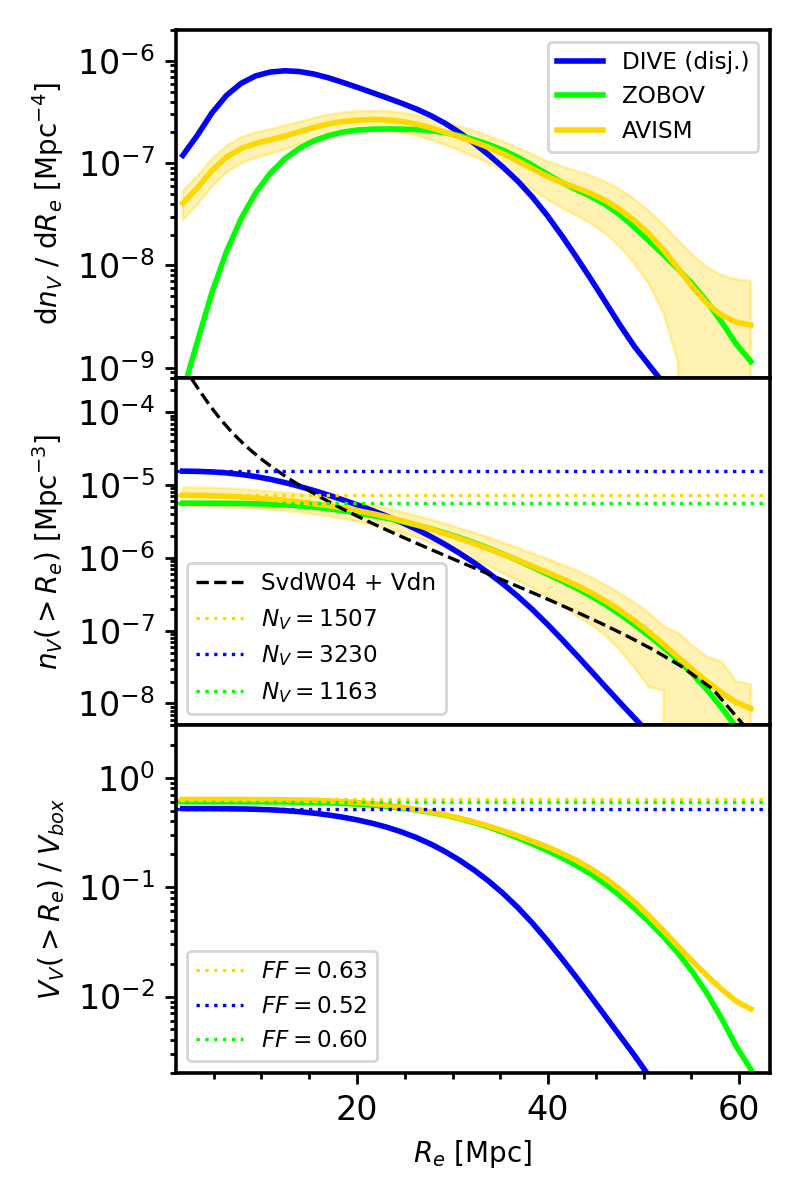}
\caption{Statistical comparison of the void distribution as found by \texttt{DIVE} (blue), \texttt{ZOBOV} (green) and \texttt{AVISM} (yellow) using the \texttt{mini-UCHUU} halos catalogue at $z =0$ as input. Top panel: void size function (VSF). The gold-shaded region represents 2$\sigma$ of the Poisson shot noise error. Middle panel: cumulative VSF with horizontal lines depicting the total void number density and the corresponding total void count ($N_V$). The black dashed line represents the best match for the theoretical SvdW+Vdn model \citep{sheth2004hierarchy, jennings2013abundance}. Bottom panel: volume filling fraction of voids above a given radius. Horizontal lines depict the total filling fraction ($FF$).}
\label{fig:void_statistics}
\end{figure}

\begin{figure}
\centering 
\includegraphics[width=0.78\linewidth]{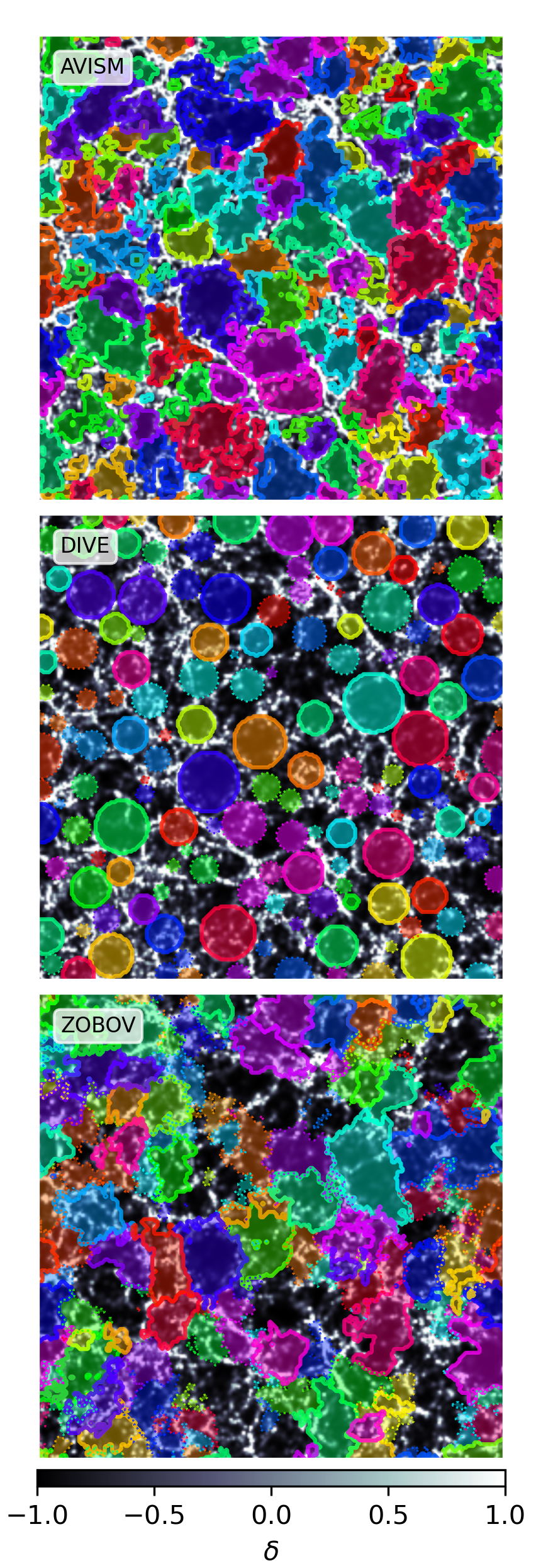}
\caption{Distribution of voids intersecting a thin slice of 400 $\mathrm{Mpc}/h$ side length through the centre of the box. Top, middle, and bottom panels show, respectively, results from \texttt{AVISM}, \texttt{DIVE}, and \texttt{ZOBOV}. Different colours are used to show void zones. Voids matching another from the reference catalogue (\texttt{AVISM} in this case) with DSC coefficient larger (smaller) than 0.4 are displayed using the same colour and continuous (dotted) lines. Voids are shown overlaying the integrated contrast density field as interpolated by \texttt{AVISM}, represented in a grey colour scale with values displayed in the colorbar below.}
\label{fig:voids_all}
\end{figure}

\begin{figure}[h!]
\centering 
\includegraphics[width=1\linewidth]{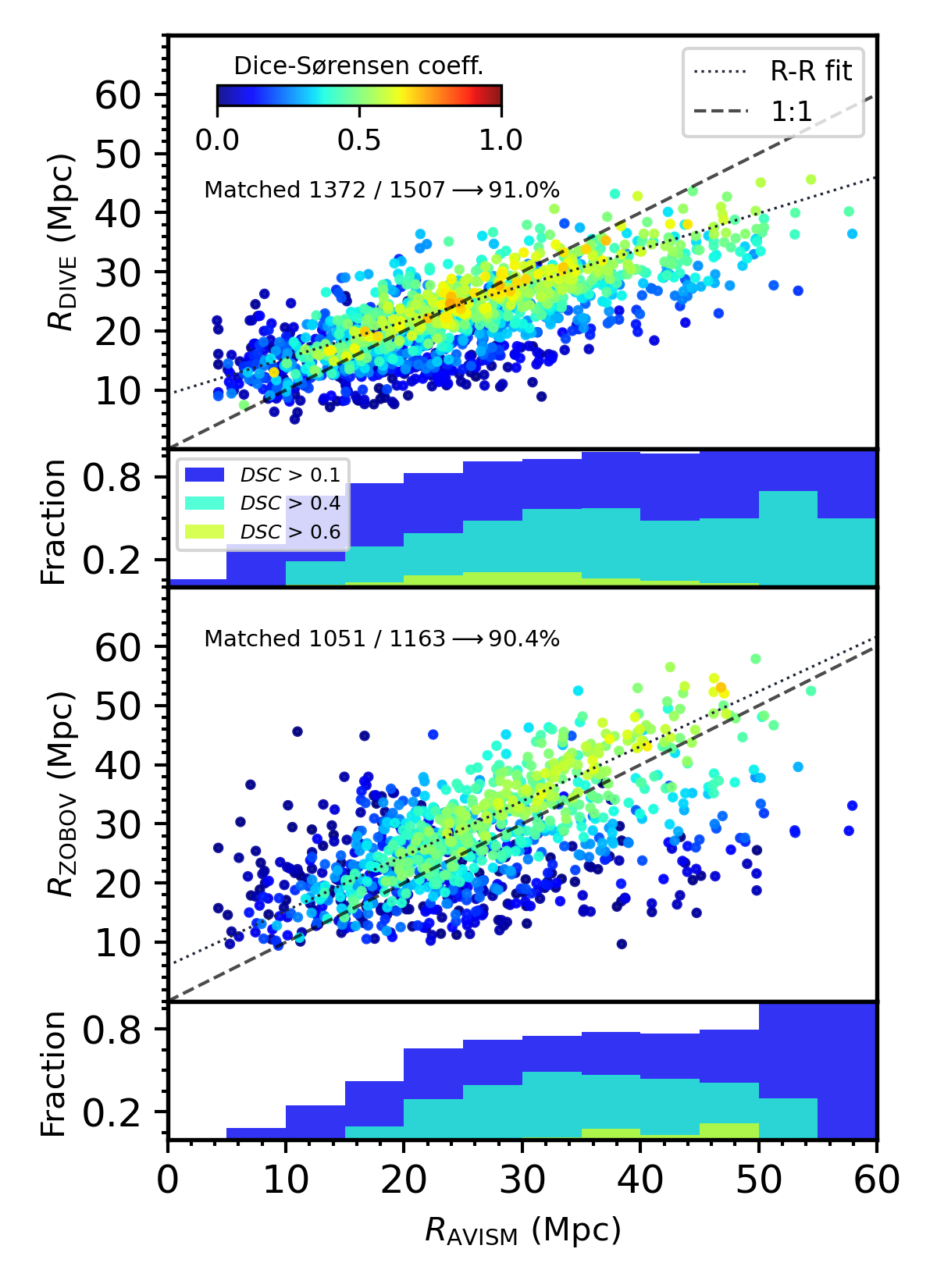}
\caption{Cross-match between \texttt{AVISM}'s voids and those found by \texttt{DIVE} (top panel) and \texttt{ZOBOV} bottom panel). For all \texttt{AVISM} voids, a point is drawn with the best match found in the other catalogues displaying, first, the colour-coded Dice-Sørensen coefficient for the match and, second, the radius of the corresponding counterpart on the vertical axis. The dashed black line shows the perfect situation in which voids matched among the void finders would have the same effective radius, whilst the dotted line displays a linear fit to the $R-R$ relation, weighted by the DSC values. The small panels below each major panel show the fraction of voids, for each radius, that have been matched with a DSC above a certain value, given by the different colours displayed in the palette. Redder (bluer) colours indicate higher (lower) DSC, meaning that the matched voids are more similar (different).
}
\label{fig:scatter_compare}
\end{figure}

\begin{figure}[h!]
\centering 
\includegraphics[width=1\linewidth]{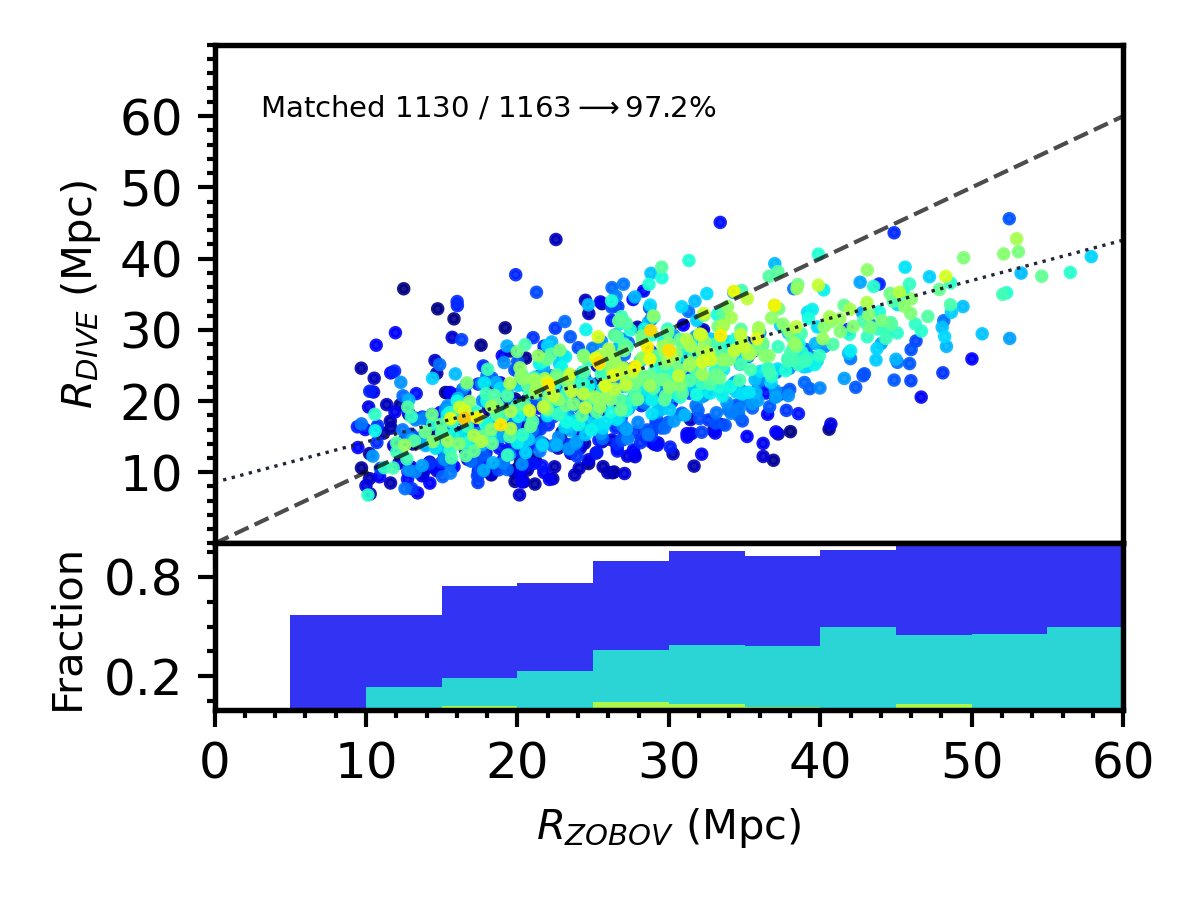}
\caption{Analogous to Fig. \ref{fig:scatter_compare}, but in this case, \texttt{ZOBOV} and \texttt{DIVE} voids have been cross-matched for comparison.}
\label{fig:scatter_compare_2}
\end{figure}

Figure \ref{fig:void_statistics} presents the outcome of the different void finders applied to the \texttt{mini-UCHUU} halos catalogue at $z =0$. In the top panel, the VSF is presented together with 2$\sigma$ of the Poisson shot noise error for the \texttt{AVISM} case. Next, in the middle panel we find the cumulative VSF with the same error and a (best) theoretical fit following the model developed in \citet{sheth2004hierarchy} (henceforward, the SvdW model) and further expanded in \citet{jennings2013abundance} (Vdn model), where the excursion-set formalism is used and voids are treated as spherical regions around density minima. In \mbox{Appendix \ref{sec:appendix_excursionset}} we provide a brief summary of the basic concepts and considerations used to compute the fit. In the bottom panel, we present the cumulative volume filling fraction as a function of radius. This last plot describes how much volume is occupied by voids larger than a given effective radius. 

\texttt{AVISM} and \texttt{ZOBOV} display similar behaviours in the size function, both in the cumulative and differential representations, with the first finding significantly more small voids. \texttt{DIVE} has the largest void population with $3230$ voids found, but it is shifted towards small sizes. None of the algorithms is able to closely follow the theoretical fit, however, \texttt{AVISM} and \texttt{ZOBOV} show a consistent trend with it for $R_e \gtrsim 15 \; \mathrm{Mpc}$ within 2$\sigma$ of the Poisson error. The deviation from the theoretical fit is mostly due to the arbitrary shapes voids can have (except for the \texttt{DIVE} case), which hugely differ from those assumed in the spherical formalism (see Appendix \ref{sec:appendix_excursionset} for more details). Moreover, the mean density contrast inside voids varies on each case and can significantly deviate from $\delta = -0.8$, hence breaking again the conditions under which the SvdW formalism is applied. Overall, the three algorithms approximately converge in terms of volume filling fraction, with \texttt{AVISM} maximising the covering (63\%). This result indicates that the three algorithms are able to detect the same total volume in voids, whereas this total volume is distributed in void catalogues with different ranges of sizes and shapes.

For the sake of a visual comparison, Fig.~\ref{fig:voids_all} displays a thin slice through the centre of the box, showing those voids intersecting the considered slice as found by each method. Voids are presented overlaying the projected contrast density field as interpolated by \texttt{AVISM}. Furthermore, so as to get a more detailed visual inspection of the three samples, we try to match the individual voids produced by the three codes inside the slice (not the entire input box). To do so, we use the DSC coefficient as defined in Appendix~\ref{sec:appendix_metric} as a metric to measure the similarity of voids. In a similar manner, as in Sect.~\ref{sec:TNG_tracers}, and taking \texttt{AVISM} result as the reference one, voids in the middle and bottom panels in Fig.~\ref{fig:voids_all} matching a void from this reference catalogue with a DSC value larger than 0.4  are displayed with a continuous line with the same colour as in the top panel. Similarly, counterparts with a DSC rate smaller than 0.4 are plotted with the same colour but using dotted lines, indicating a lower agreement. The three algorithms successfully identify most major voids in the intersecting slice. However, a region classified as a single void by one method may be divided into two distinct regions by another. Additionally, in some cases, a zone where one algorithm fails to detect a void is successfully identified by another. Hence, although they are statistically similar, \texttt{ZOBOV} and \texttt{AVISM} can find different void shapes, sizes and centres. It is also interesting how the centres and sizes found by \texttt{DIVE} and \texttt{AVISM} coincide in many cases, in spite of having such divergent methodologies to identify voids.

In order to get a more quantitative comparison among finders, we now calculate the DSC (see Appendix~\ref{sec:appendix_metric}) of all voids inside \texttt{AVISM}'s catalogue against the other two. This time, the cross-match has been carried out with all voids inside the input box, and we do not allow for duplicates, that is, a match cannot be shared by two different voids from the same catalogue. \mbox{Fig. ~\ref{fig:scatter_compare}} displays the cross-match between \texttt{AVISM}'s voids and those found by \texttt{DIVE} and \texttt{ZOBOV}, displaying the radius identified by the void finders against each other together with the colour-coded DSC corresponding to each match. The small panels below the major ones show the fraction of voids, at each radius, that have a match with a DSC larger than a certain value given by the colour palette. Redder (bluer) colours display higher (lower) agreement between the void finders. The dashed black line shows the perfect situation in which voids matched among the void finders would have the same effective radius, whilst the dotted line displays a linear fit to the $R-R$ relation, weighted by the DSC values. Strikingly, despite their very different natures, \texttt{AVISM} and \texttt{DIVE} display the best agreement when considering the volume intersection. Indeed, the fraction of \texttt{AVISM} voids intersecting with \texttt{DIVE} voids with a DSC above $0.4$ ranges from $\approx 40\%$ to $\approx 70\%$ for $R_e \gtrsim 20 \; \mathrm{Mpc}$. Also, a non-negligible fraction of $10-20\%$ voids with an overlapping index above $0.6$ can be found, especially for the middle-sized part. This indicates that both methodologies are, to some extent, placing voids in similar places with alike volumes. Regarding sizes, although the scatter is considerable, it is lower than the \texttt{AVISM} vs \texttt{ZOBOV} case; nevertheless, the $R_\texttt{AVISM}$ vs $R_\texttt{DIVE}$ fit significantly deviates from the 1:1 relation. This can be explained by the fact that \texttt{DIVE} finds, in general, smaller voids than \texttt{AVISM}.

The comparison of \texttt{AVISM} and \texttt{ZOBOV} voids shows more scatter when it comes to the size-to-size correlation, although the $R_\texttt{AVISM}$ vs $R_\texttt{ZOBOV}$ fit almost lies on top of the ideal 1:1 correspondence, since both approaches yield a similar size distribution. The DSCs are generally worse than the cross-match of \texttt{AVISM} and \texttt{DIVE} catalogues. For $R_e \gtrsim 20\; \mathrm{Mpc} $, the fraction of voids intersecting with a DSC above $0.4$ ranges from  $30\%$ to $50\%$, approximately, with some matches fulfilling $\text{DSC} > 0.6$ at large radii. One would expect \texttt{AVISM} and \texttt{ZOBOV} to have a better match, as they both allow arbitrary void shapes and yield a similar VSF. A plausible explanation for this divergence is their dissimilar definitions of voids, which, especially for the smaller ones, can return them in very different places and sizes. In fact, as can be seen in Fig. \ref{fig:voids_all}, while \texttt{ZOBOV} identifies voids that are excluded by \texttt{AVISM} due to their high densities, it struggles to find voids in very underdense regions, possibly due to the small number of particles (numerical tracers), whereas the other two void finders successfully identify them.

For the sake of completeness, Fig.~\ref{fig:scatter_compare_2} provides a cross-match of the \texttt{ZOBOV} and \texttt{DIVE} void catalogues. From all the comparisons, this is the best in terms of raw matching, as 97\% of \texttt{ZOBOV}'s voids are matched by \texttt{DIVE}'s. Nevertheless, the quality of these is not as high as the \texttt{AVISM} vs \texttt{DIVE} case: the fraction of voids intersecting with DSC above $0.6$ is never higher than $5-10\%$, and those intersecting with DSC $ >0.4$ are never higher than a $\approx 50\%$ fraction. Concerning sizes, a similar correlation to the $R_\texttt{DIVE}$ vs $R_\texttt{AVISM}$ fit is obtained, with similar scatter and slope. This is, again, due to the fact that \texttt{DIVE}'s voids are smaller than those identified by the other two approaches.

Finally, it can be seen that, in all the comparisons we have carried out, the agreement between the void finders maximises at larger void sizes and starts to decline at lower radii. This can be explained by the fact that, unlike the big ones, small voids are hugely affected by Poisson noise, as the number of resolution elements defining them is poor and, thus, little changes in the sampling or methodology can yield very different results (centre placement, size, etc.).

\begin{figure*}[h!]
\centering 
\includegraphics[width=0.9\linewidth]{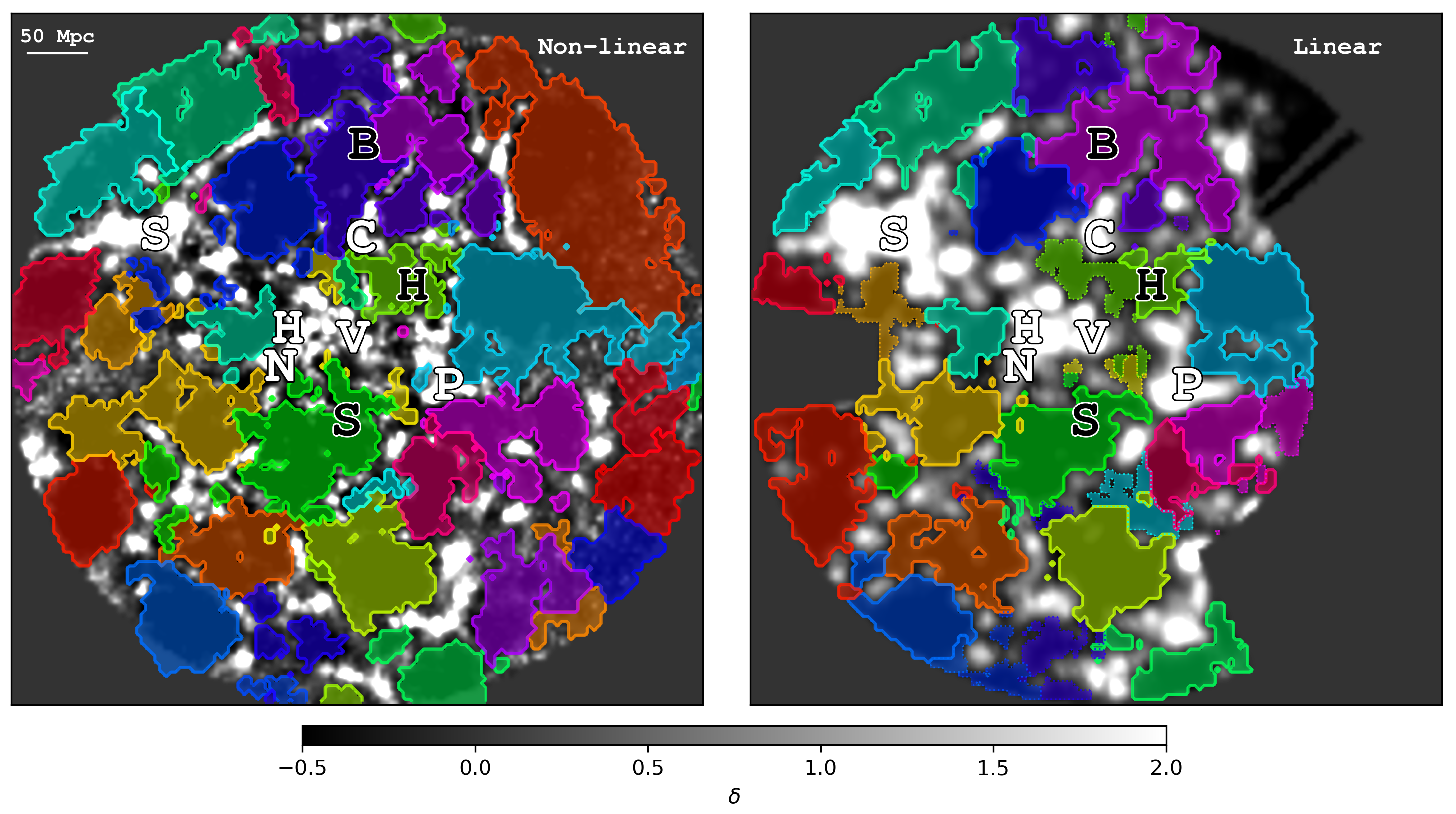}
    \caption{Slice through the centre of the \texttt{2M++} galaxy survey with all voids intersecting it as identified by \texttt{AVISM} when applied to the \citet{mcalpine2025manticore} (non-linear) and \citet{carrick2015cosmological} (linear) density and velocity reconstructions. We use the supergalactic coordinate system (SGX, SGY, SGZ) defined by \citet{de1991book}. The slice is $ \approx 8 \, \mathrm{Mpc}$ deep and contains the supergalactic plane (SGZ = 0). The void-finding procedure was restricted to the inner $R \leq 200/h \; \mathrm{Mpc}$ in both cases. Near cosmological structures are highlighted with capital letters: the Virgo (V), Hydra-Centaurus (H), Norma (N), Shapley (S) and Perseus (P) clusters together with the Sculptor (S), Hercules (H) and Boötes (B) voids \citep{tully2019cosmicflows}. 
     Voids from the linear catalogue matching another from the non-linear with DSC coefficient larger (smaller) than 0.4 are displayed using the same colour and continuous (dotted) lines. Gray scale displays integrated density contrast.}
\label{fig:manticoreVSlinear}
\end{figure*}

\section{Application to survey data}
\label{s:survey}

One of the main goals of this project is to design a void finder algorithm which can be applied either to cosmological simulation outputs or to real survey data. Whereas in the first case, the use of the void finder can be straightforward, as the density and velocity fields are generally known, the second situation could be more complex. In this Section, we discuss how these fields could be estimated in order to \texttt{AVISM} be applied to galaxy catalogue surveys.

The estimation of the density and velocity fields requires a careful treatment due to inherent problems like sample completeness, galaxy bias or RSDs. Therefore, the problem of reconstructing such fields in galaxy surveys is an open tough issue that involves the work of many groups nowadays. Thus, for \texttt{AVISM} to be successfully applied to observational data, it is necessary to transform the raw galaxy distribution into the density and velocity fields evaluated onto a cubic grid considering all the pertinent corrections. Especially important is the case of the velocity field, whose use to identify voids is a distinguishing feature of \texttt{AVISM}. As a consequence of this, the application of \texttt{AVISM} to observational data requires a pre-processing step, and the use of complementary tools to reconstruct the density and velocity fields is compulsory, being the capabilities of such field reconstruction procedure crucial on the void finder performance. 

A first approach that one could think of would be to create a continuous density field using the galaxies as mass particles conveniently smoothed onto a grid and corrected from completeness, bias and RSDs. Later, the use of the linear approximation (Eq.~\ref{div_estimation}) would provide us with the velocity divergence. This would be a misleading strategy, as no new information would be introduced besides the one provided by the density field and, therefore, the velocity divergence condition would be superfluous.

As previously mentioned, the reconstruction of the density and velocity fields associated with observational data beyond the linear regime is an extremely difficult task. Nevertheless, several options have recently produced huge advances in the topic. Let us describe briefly some of these new options. The first one is the approach based on Bayesian inference frameworks like \texttt{BORG} \citep{jasche2019physical} or \texttt{COSMIC BIRTH} \citep{kitaura2021cosmic}. These methods hinge on the basic idea of producing constrained initial conditions that conveniently evolved in a suite of numerical simulations, lead to matter distributions at $z\sim 0$ compatible with the considered observational data. Thus, non-linear density and velocity fields are obtained. The second family of methods uses neural networks (NNs) that have been trained working with several simulation datasets. Once the NNs are trained, they are properly fed with the observational data, giving the non-linear density and velocity fields as the output \citep{wu2021cosmic,lilow2024neural}.

We applied  \texttt{AVISM} to the \texttt{2M++} survey \citep{lavaux20112M++}, which is a superset of the all-sky \texttt{2MRS} survey \citep{huchra20122mass}. We use two methods to reconstruct the density and velocity fields. The first one uses the methodology described in \cite{carrick2015cosmological}\footnote{\url{http://cosmicflows.iap.fr}}
to produce a linear estimate of those fields, and therefore, as discussed before, not introducing additional information concerning the velocity field beyond the one given by the density reconstruction. A similar approach is the one used by the CORAS code 
\citep[COnstrained Realizations from All-sky Surveys;][]{lilow2021constrained} to analyse the \texttt{2MRS} survey.
The second method uses data from \texttt{Manticore-Local} \citep{mcalpine2025manticore}, where a suite of $N$-body simulations were carried out starting from constrained initial conditions produced by the \texttt{BORG} code \citep{jasche2019physical} compatible with the data from the \texttt{2M++} galaxy survey. This methodology allows for obtaining a set of realisations with the fully non-linear density and velocity fields. The void finder is applied to the averaged fields considering the whole suite of realisations.

The outcome of this test are two all-sky void catalogues within a radius of $200 \; \mathrm{Mpc}/h$. Fig.~\ref{fig:manticoreVSlinear} shows two slices through the centre of the survey with all voids intersecting it as found by \texttt{AVISM} when supplied with the non-linear reconstructed density and velocity fields given by \texttt{Manticore-Local} (left panel) and the linear fields obtained by \cite{carrick2015cosmological} (right panel), respectively. To compare the two catalogues obtained, we perform a similar comparison as in Sect. \ref{s:comparison} with the different void finders. We cross-matched both \texttt{2M++} catalogues to quantify to what extent the samples correlate. We find that $33\%$ of voids agree with a DSC above 0.4 and only $12\%$ show a DSC above 0.6, meaning that the non-linear features introduced in the \citet{mcalpine2025manticore} realisations, and not present in the \citet{carrick2015cosmological} linear reconstructions,  play an essential role in order for \texttt{AVISM} to properly identify voids. The linear reconstruction has been downgraded to ensure the same spatial resolution than in the \texttt{Manticore-Local} data, which is $\approx 3.9$ Mpc. Both void catalogues have been obtained using the same set of thresholds and parameters in \texttt{AVISM}.

Let us note that, although errors in the velocity field could be high, their impact would not be a critical issue for \texttt{AVISM}, as voids are identified (by default) under the assumption of being expanding regions with positive velocity divergence, that is,  \texttt{AVISM} is only interested in the divergence sign, as negative divergence would mark non-void regions. This is a crucial feature of our void finder, that relieves the impact of large errors in the velocity divergence, being the void finder able to reasonably recover the distribution of voids with their complex 3D shapes as long as the velocity divergence sign is correct.

We wish to clearly state that in the case of observational data that has not been preprocessed to provide either the density or the velocity information, \texttt{AVISM} can not be directly applied, and must act collaboratively with some external tool able to reconstruct the required fields. However, rather than a problem, this is a new possibility of collaboration and integration with the already mentioned new tools.

\section{Summary and conclusions}
\label{s:conclusions}

In this paper, we have presented \texttt{AVISM}, a novel void finder algorithm designed to identify cosmic voids within large-scale structure datasets. The algorithm has been thoroughly tested and validated across a wide range of scenarios, including mock void catalogues, the full output of the \texttt{TNG300-2} cosmological simulation (dark matter, gas, halos and galaxies), a dark matter halos catalogue from the mini-UCHUU simulation and real galaxy survey data. Moreover, an extensive comparison has been carried out with two other state-of-the-art void identification algorithms, namely the \texttt{DIVE} and \texttt{ZOBOV} codes. Our results demonstrate the robustness and versatility of the method in identifying voids within the LSS of the Universe, providing valuable insights for their distribution and properties.

\texttt{AVISM}'s performance has also been rigorously evaluated in terms of computational efficiency and scalability. We have tested its behaviour with varying input sizes, including the number of particles and the resolution of the auxiliary grid, and analysed its CPU scaling and efficiency. These tests confirm that the algorithm is capable of handling efficiently large datasets, both in terms of memory management and wall time, making it a practical tool for analysing current and future cosmological data, including simulation outputs --such as dark matter halos and particle information-- as well as observational data from galaxy surveys.

The idea underpinning \texttt{AVISM} is that voids are expanding, low-density, large structures. Here, we provide a summary of its methodology, performance, and applications:

\begin{enumerate}
\item \texttt{AVISM}'s void-finding methodology:
\vspace{0.2cm}

The code finds voids by means of a uniform auxiliary grid, in which density gradients and velocity divergences are computed. It is able to handle both Eulerian and Lagrangian data. In the first case, the creation of the auxiliary grid is straightforward using the original data structure. In the second one, continuous density and velocity fields are computed on the grid by means of an interpolation method similar to the SPH approach. When the original data does not include velocities (a common situation in observational data), several strategies are outlined to address this issue. Although information on velocities is needed to compute velocity divergences, the accuracy of how these velocities are reconstructed is a minor issue as far as the correct sign and ordering of the velocity divergence is caught.
    
With density contrast and peculiar velocity divergence computed on the auxiliary grid, these quantities are used, first, to label cells as candidates for void centres when they satisfy two basic conditions ($\delta < \delta_1$ and $\nabla \cdot \mathbf{v} > \nabla \cdot \mathbf{v}_\text{th}$) and, later, to expand them under some conditions. In the end, this process yields a set of cubes $\{C_i\}_{i=1}^{N_C}$ covering all regions susceptible to being part of a void. After a volume-ordered merging process, a set of unstructured non-overlapping simply connected voids $\{V_k\}_{k=1}^{N_\text{voids}}$ is obtained.

The previously described procedure can be repeated using a set of finer nested grids with higher spatial resolution. In this manner, a whole hierarchy of voids-in-voids and a deep insight into the voids' substructure can be easily achieved.

\vspace{0.5cm}
\item Tests and computational performance:
\vspace{0.2cm}

We have run \texttt{AVISM} on a mock test where a set of idealised voids were located. The exact properties of this void collection are completely known. The void finder is able to recover most quantities with errors ranging between $1\%$ and $20\%$, strongly depending on void sizes, being the smaller voids described with fewer numerical resolution elements and, therefore, presenting higher errors.

To test the versatility and robustness of the algorithm, we apply \texttt{AVISM} on the full output of the \texttt{TNG300-2} simulation from the \texttt{IllustrisTNG} suite using dark matter and gas particles, dark matter halos and galaxies as different numerical tracers. Though most voids are well identified when running the void finder on the different matter tracers, and several of them are identified as the same by the four different runs, noticeable differences arise in the void spatial distribution, as the density and velocity fields used to obtain them present different features, due to the differences in number density, positions and masses of the tracers involved. Despite that, the size distribution and, thus, the void statistics seems to be robust against high variations in the number of numerical tracers used to reconstruct the density and velocity fields (from $10^6$ to $10^9$)\footnote{Large particle numbers above $10^{10}$ can be treated without hassle or special modifications, although the allocated resources would increase, as well as computing time, that could be around 5 hours.}.

Regarding computational performance, the algorithm displays $\mathcal{O}(N_\text{cell})$ and $\mathcal{O}(N_\text{part}\log N_\text{part})$ time complexities, thus scaling well both in grid resolution and number of input particles. The code is written in \texttt{Fortran} 2008 and is parallelised using \texttt{OpenMP}, though due to the structure of some parts, the speedup when running the code in more than $\sim 32$ threads is poor. This issue will be improved in future versions.

\vspace{0.5cm}

\item Codes comparison:
\vspace{0.2cm}

We have carried out a thorough comparison between \texttt{AVISM} and two other state-of-the-art void finder algorithms, namely \texttt{DIVE} \citep{zhao2016dive} and \texttt{ZOBOV} \citep{neyrinck2008zobov, sutter2015vide}. We apply the three algorithms on the same input, consisting of a $z = 0$ snapshot from the \texttt{mini-UCHUU} $N$-body simulation \citep{ishiyama2021uchuu}, for which a dark matter halo catalogue already exists.

Due to their unstructured void-finding behaviours, \texttt{AVISM} and \texttt{ZOBOV} display similar behaviours when it comes to the VSF, with the first finding a larger population of voids at smaller radii. The spherical void finder \texttt{DIVE} obtains a void sample shifted towards smaller sizes. 

When looking  at the one-by-one match between the three codes, 
\texttt{AVISM} and \texttt{DIVE} display the best agreement in terms of volume overlapping, as their matches have the highest Dice-Sørensen coefficients, whereas in the comparison between \texttt{AVISM} and \texttt{ZOBOV}, the overlapping scores are lower, displaying a higher disagreement between the two algorithms. The size correlation cross-match between \texttt{AVISM} and \texttt{ZOBOV} voids is closer to the perfect match trend (1:1), as their distribution of sizes is similar. For completeness, a \texttt{DIVE} vs \texttt{ZOBOV} comparison is also carried out, the agreement being between the other two comparisons.
    
As a general conclusion, the three algorithms are somehow able to find similar statistical properties despite their different natures. However, although correlated, they identify different void populations (placed in different regions), since only a small fraction of voids ($20\%$ at most) have a good agreement in terms of volume overlapping ($\text{DSC} > 0.6$).
    
\vspace{0.5cm}

\item Application to real observational data:
\vspace{0.2cm}

\texttt{AVISM} can be used to search for voids in galaxy surveys. The code can internally compute the density field from a particle-like input provided by the user, with the position and mass of each galaxy, and the velocity divergence is calculated using linear theory. In this case, the velocity divergence condition does not introduce any additional information beyond the density field, leading to results that do not take advantage of the full potentialities of the code. Nevertheless, in this kind of applications, it is strongly recommended to pre-process the survey data with complementary external tools like the ones described in \cite{mcalpine2025manticore} or \cite{lilow2024neural}, which are able to produce cubic grids with the non-linear density and velocity fields.
 
As a demonstration, we have identified voids within the \texttt{2M++} survey \citep{lavaux20112M++} using data from a linear reconstruction of the density and velocity fields given by \cite{carrick2015cosmological} and the non-linear \texttt{Manticore-Local} \citep{mcalpine2025manticore} output as inputs for \texttt{AVISM}. We provide two slices aligned with the supergalactic plane (Fig.~\ref{fig:manticoreVSlinear}), with all voids intersecting it.

\end{enumerate}

\texttt{AVISM} is a public tool that could be widely used, both on simulated and observed data. A brief summary of its strengths that would justify its application to future data sets are the following: 
\begin{itemize}
    \item The code defines voids as expanding, low-density regions, and it uses geometrical and dynamical information to search for them.
    
    \item The code can handle raw simulation outputs, halo catalogues and galaxy surveys, taking into account periodic boundary conditions if the user requires it.
    
    \item No prior on the void shapes is assumed, obtaining a full 3D description of these regions.

    \item Large data volumes can be treated (i.e., more than $10^{10}$ particles) due to the code's parallelisation and optimised time complexity.
  
    \item The outcome of the void finder strongly depends on grid resolution, but weakly on numerical tracer density (galaxies, dark matter halos, ...), as particle data are interpolated onto an auxiliary grid. This sets our method apart from other existing tools.
    
    \item Although the voids found by this approach are unstructured, the complete 3D shape is defined on the auxiliary grid. This allows the user to easily distinguish which region belongs to each void and permits an easy and fast search of galaxies within voids, like in spherical void finders.

    \item The void-finding process can be applied at different levels of resolution using finer grids, leading to a list of voids-in-voids.
\end{itemize}

In conclusion, the void finder implementation presented in this work is a fast, robust and versatile choice for studying voids in the context of the large-scale structure of the Universe. Its ability to accurately and efficiently identify voids across diverse datasets makes it a powerful tool for cosmological research. Future work will focus on further expanding and optimising the algorithm, exploring its application to larger and more complex datasets, and leveraging its results to shed light on problems such as the characterisation of galaxies in voids or the constraints of cosmological parameters.

We refer the reader to Sect.~\ref{s:algorithm}, where the GitHub repository link can be found. In this public repository, we provide the user with the code's documentation, basic tools for handling simulation outputs such as the \texttt{Mini-UCHUU} halo catalogue analysed in Sect.~\ref{s:comparison} or the \texttt{TNG300-2} simulation particles, raw galaxy survey data or reconstructed grids (see Sect.~\ref{s:survey}) in order to provide a proper input for the void finder. A \texttt{Python} reader for \texttt{AVISM}'s output is also provided.

\begin{acknowledgements}

    We thank the anonymous referee for the valuable insight and constructive comments that helped improving the quality of this manuscript. We thank S. McAlpine and collaborators for providing us with the Manticore-Local data. 
     This work has been supported by the Agencia Estatal de Investigación Española (AEI; grant PID2022-138855NB-C33), by the Ministerio de Ciencia e Innovación (MCIN) within the Plan de Recuperación, Transformación y Resiliencia del Gobierno de España through the project ASFAE/2022/001, with funding from European Union NextGenerationEU (PRTR-C17.I1), and by the Generalitat Valenciana (grant PROMETEO CIPROM/2022/49).
     OM and DV acknowledge support from Universitat de València through Atracció de Talent fellowships. Simulations have been carried out using the supercomputer Lluís Vives at the Servei d'Informàtica of the Universitat de València. 
    We thank Instituto de Astrofisica de Andalucia (IAA-CSIC), Centro de Supercomputacion de Galicia (CESGA) and the Spanish academic and research network (RedIRIS) in Spain for hosting Uchuu DR1, DR2 and DR3 in the Skies \& Universes site for cosmological simulations. The Uchuu simulations were carried out on Aterui II supercomputer at Center for Computational Astrophysics, CfCA, of National Astronomical Observatory of Japan, and the K computer at the RIKEN Advanced Institute for Computational Science. The Uchuu Data Releases efforts have made use of the skun@IAA\-RedIRIS and skun6@IAA computer facilities managed by the IAA-CSIC in Spain (MICINN EU-Feder grant EQC2018\-004366\-P).
     
     This research has made use of the following open-source
     packages: NumPy \citep{harris2020array}, SciPy \citep{virtanen2020scipy}, Matplotlib
     \citep{Hunter:2007}, and CAMB \citep{lewis2011camb}.
     
\end{acknowledgements}

\bibliographystyle{aa}
\bibliography{aa54513-25}
\vspace{1 cm}

\begin{appendix}
\section{Mock test voids }
\label{sec:appendix_test}

First of all, we randomly assign a void effective radius $R_e$ according to an exponential law $P(R_e) = \exp(-aR_e)$, where $P(R_e)$ is the probability density of obtaining a void with radius $R_e$, and \mbox{$a>0$} is a constant. Then, given the mean density within $R_e$, $\rho_e$, voids are built using rejection sampling (RS) on the PDF, which is given by the universal density profile provided by  \citet{ricciardelli2013structure}. The fraction of particles inside a given radius (CDF) is:
\begin{equation}
    \mathrm{CDF}(x) = x^{\alpha + 3} \exp(x^\beta - 1)\,\, ,
\end{equation}
where $x = r/R_e$. And, thus,
\begin{equation}
    \mathrm{PDF}(x) = x^{\alpha + 2}\exp(x^\beta - 1) \bigg(1 + \frac{\beta}{\alpha + 3}x^\beta \bigg)\,\, ,
\end{equation}

When applying RS, the $\mathrm{PDF}$ can be multiplied by a constant $K_1$ in order to restrict its values to the $[0,1]$ range for $x\in[0,1]$. Once the particle positions are assigned, we can get their velocity field, which will only depend on the radial distance to the centre:
\begin{equation}
 \mathbf{v} = \mathrm{v}(r) \hat{u}_r\,\, .
\end{equation}
We impose a velocity divergence which has to decrease linearly from the centre to the border, hence:
\begin{equation}
    \nabla \cdot \mathbf{v} = \frac{1}{r^2}\frac{d}{dr}\bigg[r^2 \mathrm{v}(r)\bigg] = K_2 \frac{\mathrm{v}_0}{R_e} \bigg(D_e + 1 - \frac{r}{R_e}\bigg)\,\, ,
\end{equation}
where $\mathrm{v}_0$ is the typical velocity of particles inside the box, $K_2$ is a constant to ensure $\mathrm{v}(R_e) = \mathrm{v}_0$, and $D_e$ is another constant to constrain $\nabla \cdot \mathbf{v}(r = R_e)$.

Solving this first-order differential equation, we get:
\begin{equation}
    \mathrm{v}(r) = K_2 \mathrm{v}_0 \frac{r}{R_e} \bigg[D_e + 1 - \frac{3}{4} \frac{r}{R_e} \bigg]\,\, ,
\end{equation}
where we impose that $\mathrm{v}(r)$ cannot diverge as $r \rightarrow 0$. Furthermore, if we impose $\mathrm{v}(R_e) = \mathrm{v}_0$ and $D_e = 1$, then $K_2 = 12/5$. With this, we ensure the velocity divergence to be maximum at the void centre, decreasing linearly in the radial direction until it reaches the border, where it is still non-zero (which would be the case for $D_e = 0$).

Once we have built our spherical void, we can randomly assign three main (perpendicular) directions $a$, $b$ and $c$ in order to shrink the sphere throughout the b and c directions (by a random amount), getting an $a \ge b \ge c$ ellipsoid. Also, we have to take into account that, since we shrink the sphere, the density $\rho_e$ will rise, and this is the reason why we choose the initial $\delta_e = -0.9$ for the spheres, to obtain $\delta_e \approx -0.8$ for the resulting ellipsoids.

\section{Theoretical void size functions}
\label{sec:appendix_excursionset}

In order to derive theoretical number functions, the most common approach adopted in the field is the \citet{sheth2004hierarchy} (SvdW) model, based on the excursion-set formalism. As stated in \citet{contarini2022euclid}, the distribution of fluctuations that become voids, i.e. the multiplicity function, is obtained by this model considering a double barrier problem: a fluctuation becomes a void at a radius $r_\mathrm{L}$ if the filtered density contrast first crosses the threshold for void formation  $\delta_V^{\mathrm{L}}$ at $r_\mathrm{L}$, without having crossed the threshold for the critical overdensity for collapse  $\delta_c^{\mathrm{L}} = 1.686$\footnote{This value is well constrained by the spherical collapse model.} at any larger scale. This multiplicity function is derived in the SvdW model for spherical fluctuations in Lagrangian space, that is, with the initial density field evolving linearly to the epoch of interest. The multiplicity function as provided by SdvW is:
\begin{equation}
    f_{\ln\sigma}(\sigma) = 2\sum_{j=1}^\infty \exp\left(-\frac{(j\pi x)^2}{2}\right) j\pi x^2\,\sin(j\pi\mathcal{D})\, ,
\end{equation}
with
\begin{equation}
\mathcal{D} = \frac{|\delta_{V}^{\rm L}|}{\delta_{c}^{\rm L} + |\delta_{V}^{\rm L}|}, \qquad x = \frac{\mathcal{D}}{|\delta_{V}^{\rm L}|}\sigma \, ,
\end{equation}
where $\sigma$ is the square root of the of the variance of the linear matter perturbations on the scale $r_\mathrm{L}$ and $\delta_V^{\mathrm{L}}$ and $\delta_c^{\mathrm{L}}$ are the density thresholds discussed above.

Putting all together, we can get the void size function (VSF) in the linear regime:
\begin{equation}
    \frac{\mathrm{d} n_L}{\mathrm{d} \ln r_\mathrm{L}} = \frac{f_{\ln \sigma}(\sigma)}{V(r_\mathrm{L})} \frac{\mathrm{d} \ln \sigma^{-1}}{\mathrm{d} \ln r_\mathrm{L}},
\end{equation}
where $r_\mathrm{L}$ is the radius of a given spherical fluctuation and \mbox{$V(r_L) = \frac{4}{3} \pi r_\mathrm{L}^3$}.

Now, we can convert the linear shell radius $r_\mathrm{L}$ to the non-linear $r$ using the evolution from the linear to the non-linear epoch:
\begin{equation}
    \frac{r_\mathrm{L}}{r} = \bigg( \frac{\rho_V}{\rho_B}\bigg)^{1/3}\, ,
\end{equation}
where $\rho_V$ is the mean density inside the void and $\rho_B$ is the matter background density of the Universe. However, note, as pointed out in \citet{jennings2013abundance}, how this evolution can make the fraction of volume occupied by voids exceed unity if we preserve the void number density, as in the original SvdW model.

To overcome this, \citet{jennings2013abundance} propose a void volume-conservative model (hereafter, Vdn model) where the void volume fraction of the Universe is set to be equal in both the linear and non-linear regimes:
\begin{equation}
    V(r)\,\mathrm{d}n = V(r_{\rm L})\,\mathrm{d}n_{\rm L}\big|_{r_{\rm L}=r_{\rm L}(r)}\, , 
\end{equation}
and this provides the final VSF:
\begin{equation}
    \frac{\text{d}n}{\text{d}\ln r} = \frac{f_{\ln\sigma}(\sigma)}{V(r)} \left. \frac{\text{d}\ln\sigma^{-1}}{\text{d}\ln r_{\mathrm{L}}} \right|_{r_{\mathrm{L}} = r_{\mathrm{L}}(r)}\, .
\end{equation}

The value of $\delta_V^{\mathrm{L}}$ depends on the non-linear to linear mapping $r_L(r)$, which in turn depends on the shape of the void and which tracer is used to define it \citep{sutter2014sparse}. Until a complete theory for the VSF that accounts for all these facts exists, $\delta_V^{\mathrm{L}}$ is left as a parameter that must be adjusted for each case. For the VSF plotted in Fig. \ref{fig:void_statistics}, we found \mbox{$\delta_V^{\mathrm{L}} = -0.5$} is the best value. Furthermore, we adopted the fixed radial scaling \mbox{$r = 1.7 \; r_\mathrm{L}$} assuming voids are spheres (and they evolve so) with average density $\rho_V = 0.2 \; \rho_B$ at the non-linear epoch.

\section{Metric to compare void catalogues}
\label{sec:appendix_metric}

The mathematical problem of finding numerical metrics able to quantify the degree of similarity among different sets of data is a long-lasting issue in all scientific disciplines \citep{jaccard1901etude}. For this work, we are interested in quantifying how similar or different void catalogues are, either obtained with different void finders or generated by the same code but using different sets of parameters or numerical tracers. In particular, our comparison approach focuses on trying to match the individual voids listed in the different catalogues. 

Let us consider two void catalogues, $\mathcal{A}$ and $\mathcal{B}$. For each void in $\mathcal{A}$, we find all voids in $\mathcal{B}$ with non-null volume intersection. Among the list of possible match candidates, we chose the void in $\mathcal{B}$ maximising a given overlapping score as the match. This numerical metric should be able to quantify how similar the intersected void volumes are. A common choice in other scientific fields\footnote{This metric is widely used in image segmentation algorithms for medical applications (e.g., \citealt{taha2015metrics}).}, and the one we opted for, is the Dice-Sørensen coefficient \citep{dice1945measures, sorensen1948method}, a statistic used for gauging the similarity of two sample sets, defined as:
\begin{equation}
    \text{DSC} = \frac{2\, |V_1\cap V_2|}{|V_1| + |V_2|}\,,
\end{equation}
where $|V_i|$ is the cardinality of set $V_i$. In this context, the sets denoted by $V$ represent voids, and cardinality means volume. In our case, $\text{DSC}$ quantifies to what extent the intersected void volumes are similar ($\text{DSC} \rightarrow 1$) or they are completely different ($\text{DSC} \rightarrow 0$). Note that the Dice-Sørensen coefficient is less restrictive than other common metrics such as the related Jaccard index \citep{jaccard1901etude}. This is a desirable feature, as we do not expect voids from different void-finding approaches to match to a great extent, due to their very dissimilar methodologies and outcomes. Throughout the manuscript, we will refer to the Dice-Sørensen coefficient as DSC, for simplicity.

\end{appendix}

\end{document}